\documentclass[a4paper,12pt]{article}
\pdfoutput=1
\usepackage{graphicx}
\usepackage{amsfonts,amsthm,amsmath,amssymb,graphicx,hyperref,color,youngtab}

\def\Tr{{\rm Tr\, }}

\newcommand{\be}{\begin{equation}}
\newcommand{\bea}{\begin{eqnarray}}

\newcommand{\ee}{\end{equation}}
\newcommand{\eea}{\end{eqnarray}}

\begin{document} 

\makeatletter
\@addtoreset{equation}{section}
\makeatother
\renewcommand{\theequation}{\thesection.\arabic{equation}}

\vspace{1.8truecm}

\vspace{15pt}


{\LARGE{ 
\centerline{\bf Integrable Subsectors from Holography } 
}}  

\vskip.5cm 

\thispagestyle{empty} 
\centerline{ {\large\bf Robert de Mello Koch$^{a,b,}$\footnote{{\tt robert@neo.phys.wits.ac.za}}, 
Minkyoo Kim$^{b,}$\footnote{{\tt minkyoo.kim@wits.ac.za}} }}

\centerline{ {\large\bf and Hendrik J.R. Van Zyl${}^{b,}$\footnote{ {\tt hjrvanzyl@gmail.com}}}}

\vspace{.4cm}
\centerline{{\it ${}^a$ School of Physics and Telecommunication Engineering},}
\centerline{{ \it South China Normal University, Guangzhou 510006, China}}

\vspace{.4cm}
\centerline{{\it ${}^b$ National Institute for Theoretical Physics,}}
\centerline{{\it School of Physics and Mandelstam Institute for Theoretical Physics,}}
\centerline{{\it University of the Witwatersrand, Wits, 2050, } }
\centerline{{\it South Africa } }

\vspace{1truecm}

\thispagestyle{empty}

\centerline{\bf ABSTRACT}

\vskip.2cm 

We consider operators in ${\cal N}=4$ super Yang-Mills theory dual to closed string states propagating on
a class of LLM geometries. 
The LLM geometries we consider are specified by a boundary condition that is a set of black rings on the LLM plane.
When projected to the LLM plane, the closed strings are polygons with all corners lying on the outer edge of a single ring.
The large $N$ limit of correlators of these operators receives contributions from non-planar diagrams even for the
leading large $N$ dynamics.
Our interest in these fluctuations is because a previous weak coupling analysis argues that the net 
effect of summing the huge set of non-planar diagrams, is a simple rescaling of the 't Hooft coupling.
We carry out some nontrivial checks of this proposal.
Using the $su(2|2)^2$ symmetry we determine the two magnon 
$S$-matrix and demonstrate that it agrees, up to two loops, with a weak coupling computation performed in the CFT.
We also compute the first finite size corrections to both the magnon and the dyonic magnon by
constructing solutions to the Nambu-Goto action that carry finite angular momentum.
These finite size computations constitute a strong coupling confirmation of the proposal.

\setcounter{page}{0}
\setcounter{tocdepth}{2}
\newpage
\tableofcontents
\setcounter{footnote}{0}
\linespread{1.1}
\parskip 4pt

{}~
{}~

\section{Introduction}

There has been dramatic progress in our understanding of $\mathcal{N} = 4$ super Yang-Mills theory, motivated
primarily by the AdS/CFT correspondence \cite{AdSCFT,Witten,Gubser}.
In particular, the dynamics of the subspace of the CFT Hilbert space, spanned by operators with bare dimension $J$
with ${J^2\over N}\ll 1$ is integrable in the large $N$ limit\cite{Minahan:2002ve,Beisert:2010jr}.
The discovery of this integrability has enabled an exact computation of the spectrum of anomalous dimensions
of the CFT\cite{Gromov:2013pga}.
Being exact, these results are correct even in the strong coupling limit of the theory.

Given this remarkable progress, one would like to know if there are other sectors of the theory, whose dynamics is also
integrable.
In the article \cite{Koch:2016jnm} a new integrable sector was proposed and weak coupling evidence for the proposal 
was given.
The subsector consists of small deformations about an LLM geometry\cite{LLM}.
The geometry is labeled by a boundary condition that corresponds to a set of black rings on the LLM plane.
The fluctuations propagating on this geometry are closed strings, which when projected to the LLM plane, are polygons 
with all corners lying on the outer edge of a single ring.
The operators corresponding to an LLM background with a closed string excitation have a bare dimension of order $N^2$.
Consequently we describe these operators using the adjective ``heavy''.
The large $N$ limit of the correlators of these heavy operators are not correctly captured by summing only planar diagrams: 
non-planar diagrams must be included to capture the large $N$ dynamics.
The weak coupling analysis of \cite{Koch:2016jnm} shows that the net effect of summing this huge class of 
diagrams is a simple rescaling of the 't Hooft coupling.
This is an interesting result, since it immediately predicts the anomalous dimensions of these operators in terms of the
corresponding dimensions computed in the planar limit.
It also implies that the dynamics of this subsector is integrable\footnote{For independent discussions of the possibility of integrability at the non-planar level see \cite{Bargheer:2017nne,Eden:2017ozn}.}.
Our main motivation in this article is to provide some nontrivial checks of the proposal of \cite{Koch:2016jnm}.

The proposal was checked in \cite{Koch:2016jnm} by computing the one loop anomalous dimensions and then using 
the $su(2|2)^2$ symmetry\cite{Beisert,Koch:2015pga} to propose an exact dimension.
The results agree with the strong coupling prediction obtained from string theory\cite{Koch:2016jnm}.
Further, the $su(2|2)^2$ symmetry can be used to determine the two magnon $S$-matrix\cite{Beisert,Koch:2015pga} 
and this too can be tested in perturbation theory.
We will test this $S$-matrix against a weak coupling computation performed in the CFT, up to two loops.
Our comparison always tests ratios of scattering amplitudes so that we never need the overall pase of the $S$-matrix.
We find complete agreement between the weak coupling CFT results and the expansion of the exact two
magnon $S$-matrix.

The skeptical reader will not be convinced: indeed, all quantities compared so far are completely determined by the
$su(2|2)^2$ symmetry, so one might wonder how strict these tests are.
It is well known that the $su(2|2)^2$ symmetry does {\it not} determine the overall phase of the $S$-matrix.
In this article we test this phase.
Specifically, we compute the first finite size corrections to both the magnon and the dyonic magnon.
These corrections when computed following L\"uscher (see for example \cite{Janik}) are sensitive to the overall 
phase of the $S$-matrix.
We will follow \cite{FiniteSize} and construct solutions to the Nambu-Goto action that carry finite angular momentum.
We are then able to compute the finite size correction at strong coupling.
Remarkably these corrections are given by simply rescaling the 't Hooft coupling (after proper translation to the string 
theory description) in the AdS$_5\times$S$^5$ result.
This is a non-trivial strong coupling test that the proposal of \cite{Koch:2016jnm} passes.

In the next section we will review the construction of the CFT operators belonging to our subsector.
We also describe the action of the dilatation operator. 
Section 3 explains the determination and some weak coupling tests of the two magnon $S$-matrix.
In section 4 we compute finite size corrections for magnons and dyonic magnons.
Our conclusions are presented in section 5.

\section{CFT Analysis}

The usual lore of the large $N$ expansion simply does not apply when we consider operators with a bare dimension
that scales as $N^2$\cite{Balasubramanian:2001nh,Berenstein:2004kk,LLM,CJR}: the pertubation expansion is not
organized by the genus of ribbon graphs being summed and the large $N$ limit is not captured by summing planar
diagrams.
Fortunately, at least in the free theory, representation theory can be used to provide a basis for the local operators of
the theory and to compute the sum of the complete set of ribbon graphs\cite{Balasubramanian:2004nb,deMelloKoch:2007rqf,Kimura:2007wy,Brown:2007xh,Bhattacharyya:2008rb,Brown:2008ij,Kimura:2008ac,Kimura:2012hp}.
We will focus on the restricted Schur polynomial basis.
As we will see, these operators provide a very natural description of closed strings propagating on an LLM geometry.
When loop corrections are considered, operators in the representation theory basis mix only 
weakly\cite{deMelloKoch:2007nbd,Bekker:2007ea,Brown:2008rs,Koch:2010gp,DeComarmond:2010ie,Carlson:2011hy,Koch:2011hb,deMelloKoch:2011ci,deMelloKoch:2012ck}.
This is useful since our goal is to diagonalize the dilatation operator.

\subsection{Operators dual to LLM geometries}

The LLM geometries are regular $\frac{1}{2}$-BPS geometries that are asymptotically $AdS_5 \times S^5$.
These geometries enjoy an $R \times SO(4) \times SO(4)$ symmetry \cite{LLM}. 
The general LLM geometry is described by the metric $(i,j=1,2)$
\bea
ds^2 = -h^{-2}(dt + V_i dx^i)^2 + h^2(dy^2 + dx^i dx^i) + y e^{G} d\Omega_3 + y e^{-G} d\tilde{\Omega}_3 \label{LLMgeo}, 
\eea
where
\bea
\begin{array}{cc} h^{-2} = 2 y \cosh(G),  & z = \frac{1}{2} \tanh(G) \\ y\partial_y V_i = \epsilon_{ij} \partial_j z, & y(\partial_i V_j - \partial_j V_i) = \epsilon_{ij} \partial_y z.   \end{array}
\eea
The metric is completely determined by a single function $z$ which depends on the three coordinates $y, x^1$ and $x^2$.  
It is obtained by solving Laplace's equation
\bea
\partial_i \partial_i z + y \partial_y \frac{\partial_y z}{y} = 0.  
\eea
To obtain a regular geometry, the boundary conditions for $z$ on the plane $y=0$ (which we will refer to as the LLM plane)
must be chosen carefully.
Regularity requires that $z = \pm \frac{1}{2}$ on the LLM plane.
Any given boundary condition can be represented graphically by coloring the LLM plane black when $(z = -\frac{1}{2})$ 
and white when $(z = \frac{1}{2})$ \cite{LLM}. 

We will focus on geometries given by concentric black annuli on the LLM plane.
Each possible supergravity geometry is dual to a Schur polynomial in the CFT\cite{CJR}.
The Schur polynomial $\chi_R(Z)$ is constructed using a single complex adjoint scalar $Z$.
Consequently the ${\cal R}$ charge $J$ and dimension $\Delta$ of this operator are related as $\Delta =J$ which
is the BPS condition.
There is a concrete, explicit map between the Young diagram $R$ labeling the Schur polynomials and the coloring of the
LLM plane.
Each Schur polynomial corresponds to a coloring of concentric annuli.
Vertical edges in $R$ correspond to black annuli and horizontal edges to white annuli.   
The area of each black ring is proportional to the number of rows in the corresponding a vertical edge, while the area of 
each white ring is proportional to the number of columns in the corresponding horizontal edge.  
An example of the map between LLM plane colorings and Young diagrams is illustrated in Fig.  \ref{ring} below.
\begin{figure}[ht]%
\begin{center}
\includegraphics[width=0.80\textwidth]{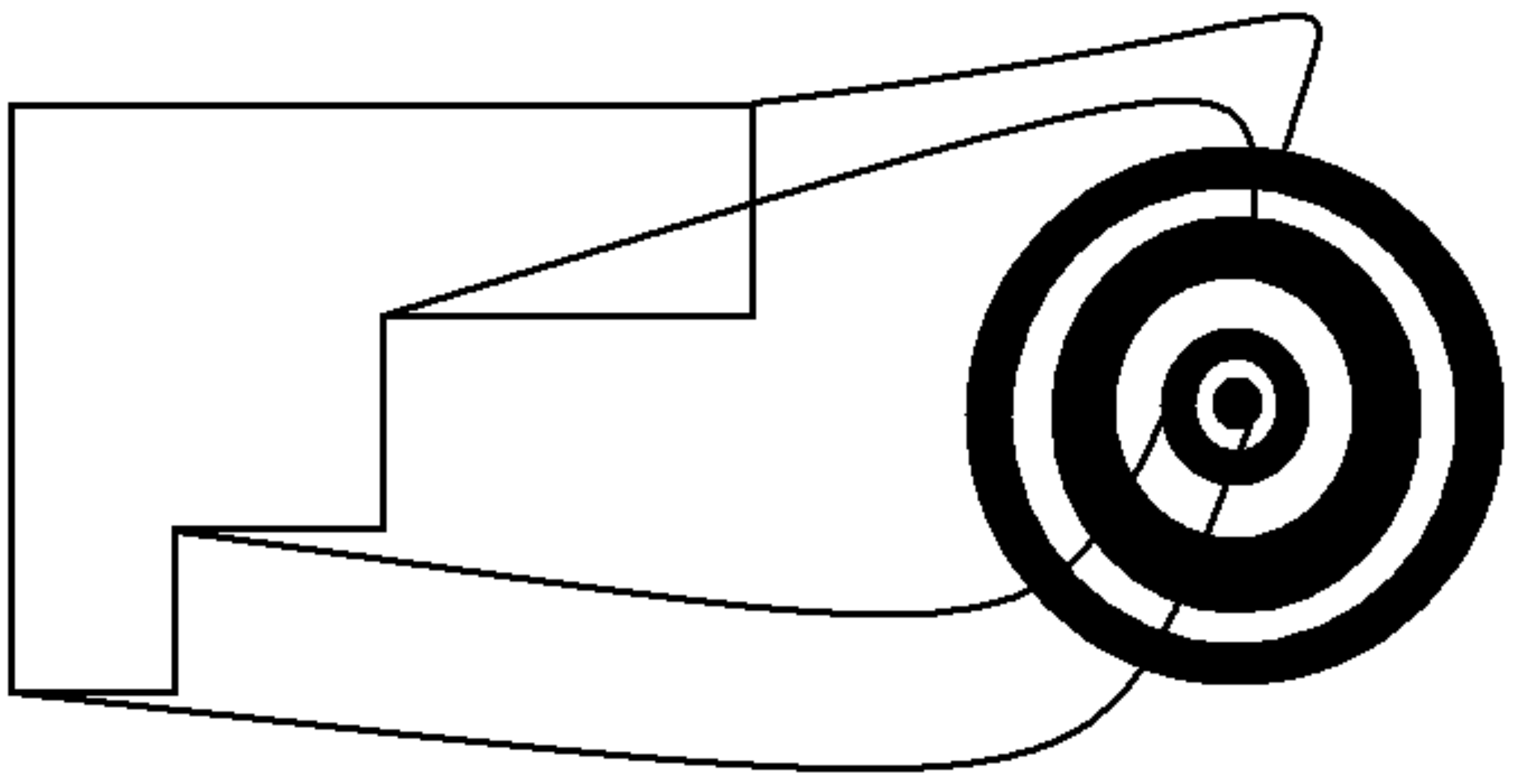}
\caption{An example of the map between Young diagrams and LLM boundary conditions.}%
\label{ring}%
\end{center}
\end{figure}
For a set of rings having a total of $E$ edges with radii $R_l$ $l=1,2,...,E$ the geometry is determined by the functions
$$ z = \sum_{l=1}^E {(-1)^{E-l}\over 2}\left( {r^2+y^2-R_l^2\over\sqrt{(r^2+y^2+R_l^2)^2-4r^2 R_l^2}}\right),$$
$$ V_{\phi}(x^1,x^2,y)=\sum_{l=1}^E {(-1)^{E-l+1}\over 2}\left( {r^2+y^2+R_l^2\over\sqrt{(r^2+y^2+R_l^2)^2-4r^2 R_l^2}}-1\right).$$
Finally, given the concrete map between the Young diagram and the LLM boundary condition, when we want to talk about
an excitation localized on the LLM plane at the edge of a given annulus we will simply talk about an excitation localized at a
corner of the Young diagram.

\subsection{Closed strings propagating on LLM geometries}

In \cite{Koch:2016jnm} operators representing excitations localised at a single corner of a Young diagram were constructed.
These operators are dual to LLM geometries excited by strings localized at the outer edge of the corresponding annulus.  
In this section we briefly review the construction of \cite{Koch:2016jnm} and then extend it to the description of closed string
states with a worldsheet that visits the edges of multiple rings.
The conjecture we want to test concerns strings localized at a single edge, but we discuss the general case for completeness.

The restricted Schur polynomials are a basis for the local operators of the theory.
Consequently, an arbitrary operator $O$ can be written as a linear combination of restricted Schur polynomials as follows 
\bea
O = \sum_{R,\{ r\}, \alpha} a_{R,\{ r\}, \alpha} \chi_{R,\{ r\}, \alpha}(Z,Y,X,\cdots)\label{CS}
\eea
The operator dual to a closed string state is a single trace operator.
Focus on a single trace operator $O$, given by the trace of a product of fields of the ${\cal N}=4$ super Yang-Mills theory.
These fields may be adjoint scalars, adjoint fermions or covariant derivatives of these fields\cite{deMelloKoch:2011vn,Koch:2012sf}.
If we use a total of $k$ species of fields then $\{ r\}$ is a collection of $k$ Young diagrams, one for each species.
If we use $n_i$ fields of species $i$ the corresponding Young diagram $r_i$ has $n_i$ boxes.
The Young diagram $r_1$ corresponds to the $Z$ field.
The Young diagram $R$ has $n_1+n_2+...+n_k$ boxes.
The additional labels contained in $\alpha$ are discrete labels distinguishing operators that carry the same $R,\{ r\}$ labels.

We will, as usual, think of the single trace operator as a one dimensional lattice.
All fields which are not $Z$s are impurities and we record their position on the lattice.
Thus, the operator $O(\{ l_1=1, l_2=4\})$ corresponds to a trace $\Tr (YZ^2YZ^{J-2})$.

Suppose now that we consider a closed string excitation of an LLM geometry, attached to a particular corner
of the Young diagram $B$ describing the LLM geometry.  
The proposal of \cite{Koch:2016jnm} is that this operator is simply given by 
\bea
O_B = \sum_{R,\{ r\}, \alpha} a_{R,\{ r\}, \alpha} \chi_{R_B,\{ r_{1B},r_2,\cdots,r_k\}, \alpha}(Z)\label{CSinLLM}
\eea
The coefficients appearing in the sum (\ref{CSinLLM}) are identical to the coefficients appearing in (\ref{CS}).
Including the background implies that we have included an enormous number of extra $Z$ fields.
Consequently, the Young diagrams $R$ and $r_1$, both of which are sensitive to the number of $Z$ fields, are modified
as indicated above.
$R_B$ (or $r_B$) is obtained by attaching $R$ (or $r$) to the relevant corner of the background Young diagram $B$.

The operator $O_B$ is significantly more complicated than the single trace operator $O$. 
Indeed, it has a complicated multitrace structure and a bare dimension of order $N^2$.
Nevertheless using the representation theory approach, it is possible to compute free field correlators exactly, and to
evaluate the action of the dilatation operator. 
Our goal in the remainder of Section 2 is to sketch how this is achieved.
As we explained above, any gauge invariant operator is given by a formula of this type
\bea
   O(Z,Y,\cdots )=\sum_{L} a_L\chi_L(Z,Y,\cdots) 
\eea
where $L$ stands for the complete label of Young diagrams plus discrete labels, spelled out above.
Recall that the content of a box in row $i$ and column $j$ of a Young diagram is given by $N-i+j$.
The coefficients $c_L$ are characters or restricted characters. 
They depend only on differences between contents of boxes and hence these coefficients are the same whether 
or not we have the background.
Now, imagine that we compute a two point function
\bea
   \langle O_{B,1}(Z,Y,\cdots)O_{B,2}(Z,Y,\cdots)^\dagger\rangle
&=&\sum_{L_1, L_2} a^1_{L_1}a^2_{L_2}
\langle \chi_{B,L_1}(Z,Y,\cdots)^\dagger \chi_{B,L_2}(Z,Y,\cdots)\rangle\cr
&=&\sum_{L} a^1_{L}a^2_{L}
\langle \chi_{B,L}(Z,Y,\cdots)^\dagger \chi_{B,L}(Z,Y,\cdots)\rangle\label{NonTrvCorr}
\eea
To move to the second line we use the fact the Schur polynomials are orthogonal.  
All of the $N$ dependence in this answer is in the values of factors of boxes in the Young diagram.
The factor of any box in the diagram is given by $N$ plus the content of the box.
When there is no background these factors are all $N$ plus an order 1 number.
In the presence of the background these factors are shifted to $N+M$ plus an order 1 number. 
$M$ is of order $N$ and it depends on the specific corner at which the excitation is localized. 
Choosing conventions so that the AdS$_5\times$S$^5$ geometry is given by a black disk on the LLM plane
of radius $1$, $\sqrt{N+M\over N}$ gives the radius at which the excitation is localized.
The computation of the correlator (\ref{NonTrvCorr}) is highly nontrivial.
It involves summing huge classes of both planar and non-planar diagrams. 
In the end however we have a rather beautiful relationship between correlators computed in the LLM background
and correlators computed in AdS$_5\times$S$^5$.
The relation is \cite{Koch:2008ah,Koch:2008cm} (see also \cite{Berenstein:2017abm,Berenstein:2017rrx,Lin:2017dnz}
for related work reaching a similar conclusion)

\bea
{\langle O_{B,1}(Z,Y,\cdots)O_{B,2}(Z,Y,\cdots)^\dagger\rangle\over \chi_B(Z)\chi^\dagger_B(Z)\rangle}
=\left( {N+M\over N}\right)^J
\langle O_{1}(Z,Y,\cdots)O_{2}(Z,Y,\cdots)^\dagger\rangle
\eea
This relation holds in the large $N$ limit and as long as the dimensions $\Delta_1$ and $\Delta_2$ of $O_1$ and $O_2$ 
obey $(\Delta_1)^2\ll N$ and $(\Delta_2)^2\ll N$.
Further, $J$ is the number of $Z$s in $O_1$ which equals the number of $Z$s in $O_2$ for a non-zero correlator.
The simplest way to summarize what is going on is simply to recognize that the complete effect of summing the huge
class of non-planar diagrams is that each $Z$ field propagator has been rescaled by $(N+M)/N$.

We will now build operators stretched between different corners of the Young diagram.  
We will distribute the operator $O$ over corners, labeled 1 and 2, of the background Young diagram.
The construction simply amounts to getting indices to contract correctly.
This is most simply achieved by writing our loop $O$ as follows
\bea
&&{\rm Tr}(Y{\partial\over \partial V}Y{\partial\over\partial W})\,\,
V^{a_1}_{b_1}(Z^{n_1})^{b_1}_{c_1}Y^{c_1}_{d_1}(Z^{n_2})^{d_1}_{e_1}
\cdots Y^{f_1}_{g_1}(Z^{n_k})^{g_1}_{a_1}\,\,
W^{a_2}_{b_2}(Z^{m_1})^{b_2}_{c_2}Y^{c_2}_{d_2}(Z^{m_2})^{d_2}_{e_2}
\cdots Y^{f_2}_{g_2}(Z^{m_q})^{g_2}_{a_2}\cr
&&\equiv
{\rm Tr}(Y{\partial\over \partial V}Y{\partial\over\partial W})\,\,
{\rm Tr}(VZ^{n_1}YZ^{n_2}\cdots YZ^{n_k})_1\,\,
{\rm Tr}(WZ^{m_1}YZ^{m_2}\cdots YZ^{m_q})_2
\eea
where subscripts $1$ and $2$ indicate which corner the fields are attached to.  
We will write the above loop as $O((n_1,n_2,\cdots,n_k)_1,(m_1,m_2,\cdots,m_q)_2)$.
We now express the single trace factors ${\rm Tr}(VZ^{n_1}YZ^{n_2}\cdots YZ^{n_k})_1$ and 
${\rm Tr}(WZ^{m_1}YZ^{m_2}\cdots YZ^{m_q})_2$ as a sum of rectricted Schur polynomials and then
attach each to a given corner using the method described above.
The separate Young diagrams of the two traces come together at different corners on the background.
We can then act with the derivatives with respect to $V$ and $W$ to finally contract all indices correctly.
The generalization to operators stretched over multiple corners is obvious.

\subsection{Dilatation operator}

The action of the dilatation operator is evaluated following \cite{Koch:2016jnm}, by breaking the problem
down into three steps:
\begin{itemize}
\item[1.] Write the usual planar computation in the trivial background, in terms of characters.
\item[2.] Write the computation in the LLM background in terms of characters.
\item[3.] Use large $N$ to develop a relationship between 1 and 2. 
\end{itemize}
The result is similar to what we found for correlators in the previous subsection: the net effect of summing the enormous
number of non-planar diagrams is a simple rescaling of the 't Hooft coupling.
The details of this evaluation are similar to the manipulations given in \cite{Koch:2016jnm}.
Rather than repeating these highly technical manipulations we simply state the results and refer the reader
to \cite{Koch:2016jnm} for details.

\begin{figure}[ht]
\begin{center}
\includegraphics[width=0.48\textwidth]{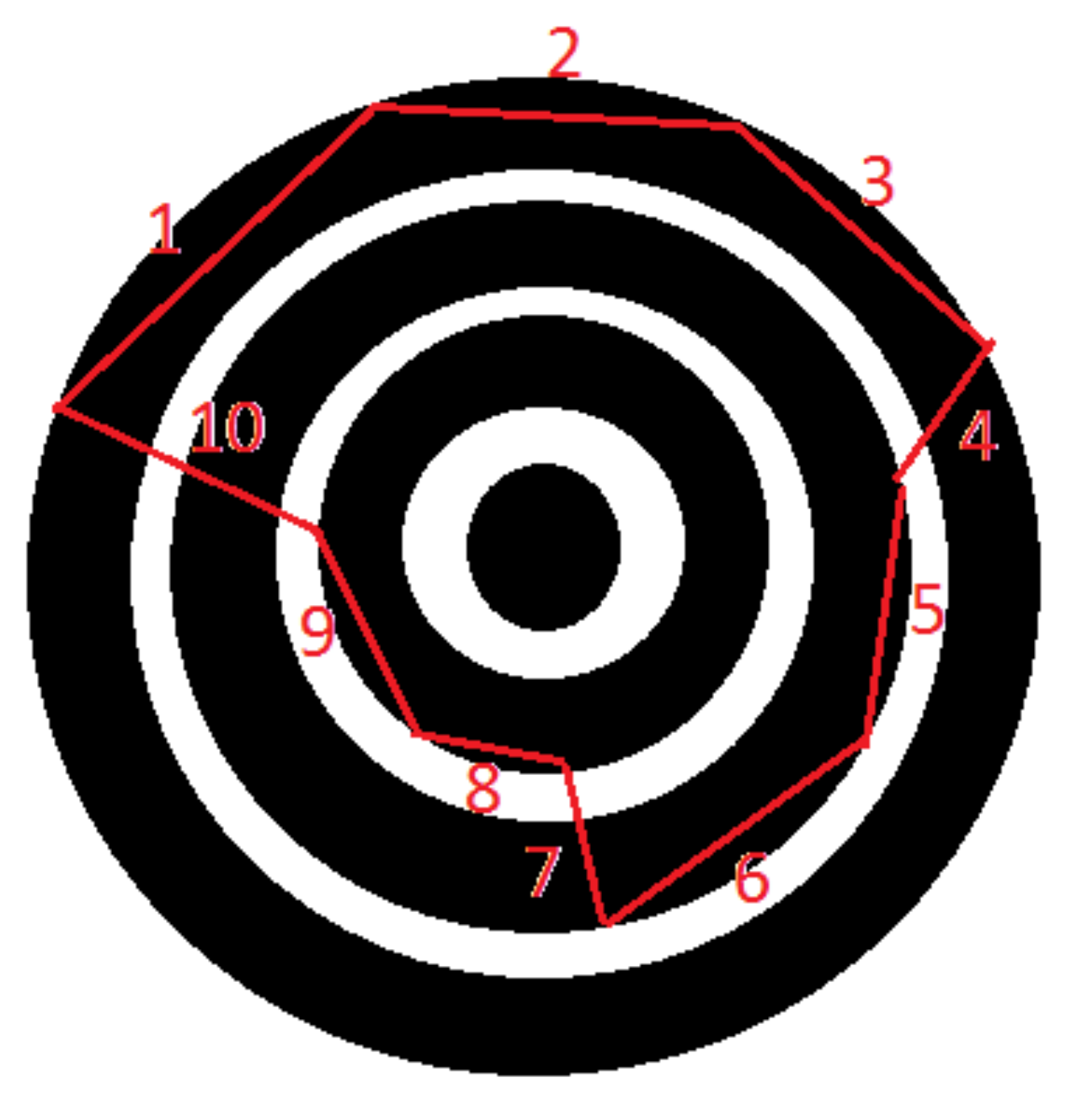}
\caption{A closed string excitation of an LLM spacetime geometry. 
The magnons on the closed string are labeled.}
\label{genmag}
\end{center}
\end{figure}

Consider the closed string excitation given in Figure \ref{genmag}.
To spell out the general result, we will give the action of the dilatation operator on a few magnons.
Denote the factors associated to the inward pointing corners of the Young diagram by $c_1>c_2>c_3>c_4$.
Recall that $\sqrt{c_i\over N}$ gives the radius squared of the outer edges of the back regions (the three rings
or central disk) on the LLM plane.
The string shown in Figure \ref{genmag} is spread over three regions, so that, in the notation of the previous
subsection we write this loop as $O_B\left((n_1,n_2,n_3,n_4)_1,(n_5,n_6,n_7)_2,(n_8,n_9,n_{10})_3\right)$.
The action of the dilatation operator on magnon 4 for example, is given by
\bea
&&DO\left((n_1,n_2,n_3,n_4)_1,(n_5,n_6,n_7)_2,(n_8,n_9,n_{10})_3\right)=\cr
&&\quad\lambda {c_1+c_2\over N}O\left((n_1,n_2,n_3,n_4)_1,(n_5,n_6,n_7)_2,(n_8,n_9,n_{10})_3\right)\cr
&&\quad-\lambda{\sqrt{c_1c_2}\over N}
\Big[O\left((n_1,n_2,n_3,n_4-1)_1,(n_5+1,n_6,n_7)_2,(n_8,n_9,n_{10})_3\right)\cr
 &&\qquad\qquad\qquad+O\left((n_1,n_2,n_3,n_4+1)_1,(n_5-1,n_6,n_7)_2,(n_8,n_9,n_{10})_3\right)\Big]
\eea
and on magnon 5 is
\bea
&&DO\left((n_1,n_2,n_3,n_4)_1,(n_5,n_6,n_7)_2,(n_8,n_9,n_{10})_3\right)=\cr
&&\quad\lambda {2 c_2\over N}O\left((n_1,n_2,n_3,n_4)_1,(n_5,n_6,n_7)_2,(n_8,n_9,n_{10})_3\right)\cr
&&\quad-\lambda{c_2\over N}
\Big[O\left((n_1,n_2,n_3,n_4)_1,(n_5+1,n_6-1,n_7)_2,(n_8,n_9,n_{10})_3\right)\cr
&&\qquad\qquad +O\left((n_1,n_2,n_3,n_4)_1,(n_5-1,n_6+1,n_7)_2,(n_8,n_9,n_{10})_3\right)\Big]
\eea
The coefficient of the diagonal term is the sum of the factors at the two ends of the magnon.
The coefficient of the hopping term is the square root of the product of the factors at the two ends of the magnon.
For a closed string with all magnon endpoints attached the edge of a single ring associated to factor $c$, this simply 
amounts to rescaling the 't Hooft coupling $\lambda\to {c\over N}\lambda$.
These are the excitations of interest to us in this study.

\begin{figure}[h]%
\begin{center}
	\centering
		\includegraphics[width=0.67\textwidth]{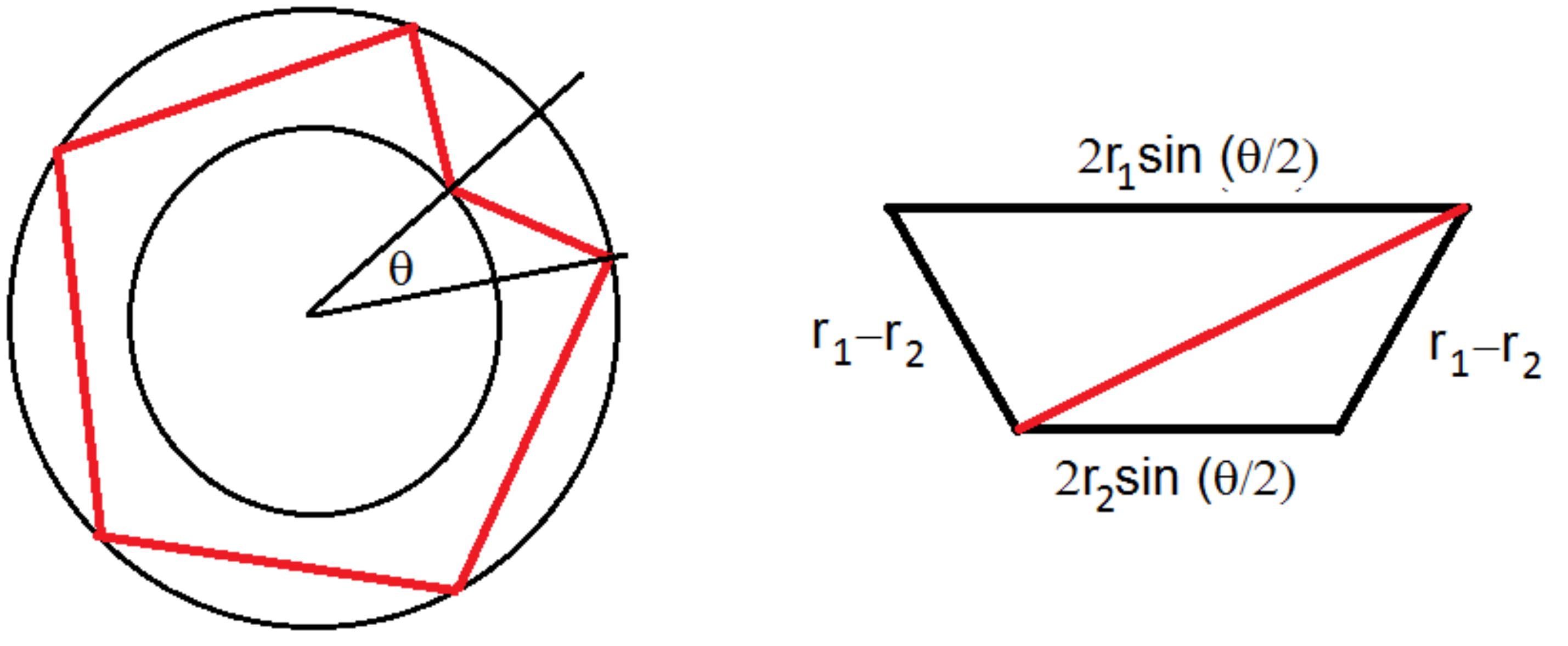}
\caption{The energy of a magnon stretching between the two radii is determined by the length of the diagonal of
the parallel trapezium shown.}%
\label{fig:energy}%
\end{center}
\end{figure}
An immediate test of this dilatation operator is to see if we obtain the expected anomalous dimension for each magnon.
Consider a magnon stretching between two different radii $r_1 > r_2$ on the LLM plane, as shown in Figure
\ref{fig:energy}.
The energy of the magnon stretching between the two radii is given in terms of the length of the diagonal of
the parallel trapezium\cite{GiantMagnons}. 
This length squared is given by Ptolemy's theorem as
\bea
   l^2 =(r_1-r_2)^2 +4r_1 r_2 \sin^2{\theta\over 2}
\eea
The energy of the magnon is given by\cite{GiantMagnons}
\bea
E=\sqrt{1+4 g^2 l^2}=1+ 2 g^2 l^2+\cdots
\eea
so that the one loop anomalous dimension is given by
\bea
\gamma = 2 g^2 ((r_1-r_2)^2 +4r_1 r_2 \sin^2{\theta\over 2})= 2 g^2 (r_1^2+r_2^2-2r_1 r_2\cos \theta)
\eea
Writing this in terms of gauge theory parameters $r_i = \sqrt{\frac{c_i}{N}}$ we have
\bea
\gamma= 2 g^2 \left({c_1\over N}+{c_2\over N}-2{\sqrt{c_1 c_2}\over N}\cos p\right) \label{Dim2Corners}
\eea
in complete harmony with our discussion above.

\section{Magnon Scattering}\label{BA}

The $S$-matrix for two magnon scattering is completely determined, up to a phase, by the $su(2|2)^2$ symmetry
that our subsector enjoys\cite{Beisert}.
In the next subsection we will obtain this two magnon scattering matrix, for the general case that the magnons
that scatter stretch between different ring edges on the LLM plane.
As a perturbative test of this result, we expand the $S$-matrix in a weak coupling expansion and demonstrate
agreement with one and two loop magnon scattering in the CFT.

A remark is in order. 
In determining the $S$-matrix up to a phase we have not made any use of integrability.
The only input is the $su(2|2)^2$ symmetry that our operators enjoy.
This symmetry is present whenever we deal with operators constructed using many $Z$s and then doped with a 
much smaller number of impurities.
This includes, for example, open strings attached to giant gravitons as well as closed string excitations of LLM geometries.

\subsection{The Exact Two-magnon S-matrix}

The parameters determining the kinematical state of a magnon stretching from radius $r_i$ to radius $r_j$ 
with momentum $p$ are
\bea
x_i^{\pm} & = & \frac{i e^{\pm i \frac{p_i}{2} }(1 + E_i)  }{2 g (r_i e^{i\frac{p_i}{2}} - r_j e^{-i \frac{p_i}{2}}  )} \nonumber \\
E_i & = & \sqrt{1 + 4 g^2 (r_i^2 + r_j^2) - 8 g^2 r_i r_j \cos(p_i) }
\eea
They satisfy the following conditions
\bea
r_i x_i^{+} +\frac{r_i}{x_i^{+}}- r_j x_{i}^{-} - \frac{r_j }{x_{i}^{-}} & = & \frac{i}{g}\cr
\frac{x_i^{+}}{x_i^{-}} & = & e^{i p_i}
\eea
The magnon 1 has its end points on rings with radii $r_0$ and $r_1$, while magnon 2 has its end points on rings 
with radii $r_0$ and $r_2$.
To specify the $su(2|2)$ representation that each magnon transforms in, following \cite{Beisert}, we specify 
parameters $a_k,b_k,c_k,d_k$, where
\bea
  Q^\alpha_a|\phi^b\rangle =a_k\delta_a^b|\psi^\alpha\rangle\,,
\qquad
  Q^\alpha_a|\psi^\beta\rangle =b_k \epsilon^{\alpha\beta}\epsilon_{ab}|\phi^b\rangle\,,
\eea
\bea
S^a_\alpha |\phi^b\rangle = c_k \epsilon_{\alpha\beta}\epsilon^{ab}|\psi^\beta\rangle\,,
\qquad
S^a_\alpha|\psi^\beta\rangle =d_k \delta^\beta_\alpha |\phi^a\rangle\,,
\eea
for the $k$th magnon.
The representation for magnon 1 is defined by
\bea
  a_1 =\sqrt{g r_0}\eta_1\,,\qquad b_1={\sqrt{g r_0}\over\eta_1} \alpha_1 f_1 \left(1-\frac{r_1}{r_0}{x^+_1\over x^-_1}\right)
\eea
\bea
c_1={\sqrt{g r_0} i\eta_1\over \alpha_1 f_1 x^+_1}\,,\qquad d_1={\sqrt{g r_0} x^+_1\over i\eta_1}
\left(1-\frac{r_1}{r_0}{x^-_1\over x^+_1}\right)
\eea
and the representation for magnon 2 by
\bea
  a_2=\sqrt{g r_0}\eta_2\,,\qquad b_2={\sqrt{g r_0}\over\eta_2} \alpha_2 f_2 
\left(1-\frac{r_2}{r_0}{x_2^+\over x_2^-}\right)
\eea
\bea
c_2={\sqrt{g r_0} i\eta_2\over \alpha_2 f_2 x_2^+}\,,\qquad 
d_2={\sqrt{g r_0} x_2^+\over i\eta_2}\left(1-\frac{r_2}{r_0}{x_2^-\over x_2^+}\right)\,.
\eea
$f_i$ is the product of $\prod_k e^{ip_k}$ for all the magnons to the left of the one considered.  
The $\alpha_i$'s are phases that position the endpoints of the magnon on the LLM plane. Denote the magnons before
scattering with 1 and 2 and after the scattering with $1'$ and $2'$.
We have
\begin{eqnarray}
\alpha_1 & = & \frac{r_0 x_1^{+} - r_1 x_1^{-}}{r_1 x_1^{+} - r_0 x_1^{-} }\alpha_2 \qquad
\alpha_{1'} =  \frac{r_0 x_{1'}^{+} - r_1 x_{1'}^{-}}{ r_1 x_{1'}^{+} - r_0 x_{1'}^{-} }\alpha_2 \qquad
\alpha_{2'} = \alpha_2.  
\end{eqnarray}
The $S$-matrix is now determined, following \cite{Beisert}, by using the magnon representation given above and 
requiring that the $S$-matrix commutes with the elements of $su(2|2)$.
We will focus on an initial state of two bosonic magnons
\bea
   S_{12}|\phi_1^a\phi_2^b\rangle = A_{12}|\phi_{2'}^{\{ a}\phi_{1'}^{b\}}\rangle +
B_{12}|\phi_{2'}^{[ a}\phi_{1'}^{b]}\rangle + {1\over 2}C_{12}\epsilon^{ab}\epsilon_{\alpha\beta}
|\psi_{2'}^\alpha\psi_{1'}^\beta \rangle
\label{frstSmat}
\eea 
The resulting S-matrix elements are given by
\begin{eqnarray}
A_{12} & = & S^0_{12} \frac{N_{A_{12}}}{D_{A_{12}}}    \nonumber \\
N_{A_{12}} & = & -x_1^{-} x_2^{-} (-x_1^{-} (x_1^{+} + x_2^{+}) + \frac{r_1}{r_0} ((x_1^{-})^2 + x_1^{+} x_2^{+})) (\frac{r_1}{r_0} x_1'^{-} - 
   x_1'^{+}) x_1'^{+} x_2'^{+} \eta_1 \eta_2 \nonumber \\
D_{A_{12}} & = & (\frac{r_1}{r_0} x_1^{-} - x_1^{+}) x_1'^{-} \left(-x_1^{-} x_2^{-} (x_1^{+} + x_2^{+}) x_1'^{+} x_2'^{+} + 
   \right. \nonumber \\ 
	& & + \left. \frac{r_1}{r_0} x_1^{+} x_2^{-} x_2^{+} (x_1^{-} x_1'^{-} + x_1'^{+} x_2'^{+}) + 
   \frac{r_2}{r_0} x_1^{+} x_2^{+} (-x_1^{+} x_2^{+} x_1'^{-} + x_1^{-} x_1'^{+} x_2'^{+}) \right) \eta_1' \eta_2' \nonumber \\
B_{12} & = & S^0_{12}\frac{N_{B_{12}}}{D_{B_{12}}} \nonumber \\
D_{B_{12}} & = & (\frac{r_1}{r_0} x_1^{-} x_2^{-} - \frac{r_2}{r_0} x_1^{+} x_2^{+}) (\frac{r_1}{r_0} x_1'^{-} x_2'^{-} - \frac{r_2}{r_0} x_1'^{+} x_2'^{+})D_{A_{12}} \nonumber \\
N_{B_{12}} & = & -x_1^{-} x_2^{-} (\frac{r_1}{r_0} x_1'^{-} - 
   x_1'^{+}) x_1'^{+} \left[4 (\frac{r_1}{r_0} x_1^{-} - x_1^{+}) x_1^{+} x_2^{+} (\frac{r_1}{r_0} x_1^{-} x_2^{-} - 
      \frac{r_2}{r_0} x_1^{+} x_2^{+}) x_1'^{-} x_2'^{-} + \right. \nonumber \\
			& & + \left. (4 \frac{r_2}{r_0} (\frac{r_1}{r_0} x_1^{-} - 
         x_1^{+}) x_1^{+} x_2^{+} (-\frac{r_1}{r_0} x_1^{-} x_2^{-} + 
         \frac{r_2}{r_0} x_1^{+} x_2^{+}) x_1'^{-} + \right. \nonumber \\
				& & + \left.(\frac{r_1}{r_0} (\frac{r_1}{r_0} x_1^{-} x_2^{-} - 2 x_1^{+} x_2^{-} + 
            \frac{r_2}{r_0} x_1^{+} x_2^{+}) (-x_1^{-} (x_1^{+} + x_2^{+}) + 
            \frac{r_1}{r_0} ((x_1^{-})^2 + x_1^{+} x_2^{+})) x_1'^{-} - \right. \nonumber \\
						& & \left. 
         - 2 (\frac{r_1}{r_0} x_1^{-} x_2^{-} - \frac{r_2}{r_0} x_1^{+} x_2^{+}) (x_1^{-} (-x_1^{+} + x_2^{+}) + 
            \frac{r_1}{r_0} ((x_1^{-})^2 - x_1^{+} x_2^{+})) x_1'^{+}) x_2'^{-}) x_2'^{+} + \right. \nonumber \\ 
						& & 
   \left. \frac{r_2}{r_0} (x_1^{-} (\frac{r_1}{r_0} x_1^{-} - x_1^{+}) (\frac{r_1}{r_0} x_1^{-} + 
         2 x_1^{+}) x_2^{-} + (3 \frac{r_2}{r_0} x_1^{-} x_1^{+} (-\frac{r_1}{r_0} x_1^{-} + x_1^{+}) + (3 \frac{r_1}{r_0} x_1^{-} - \right. \nonumber \\ 
				& & \left. - 2 x_1^{+}) (x_1^{-} - \frac{r_1}{r_0} x_1^{+}) x_2^{-}) x_2^{+} + 
      \frac{r_2}{r_0} x_1^{+} (-x_1^{-} + \frac{r_1}{r_0} x_1^{+}) (x_2^{+})^2) x_1'^{+} (x_2'^{+})^2\right] \eta_1 \eta_2   \label{ExactSMatrix}
\end{eqnarray}
where $S^0_{12}$ is an overall phase of the S-matrix, not determined by the symmetry argument. 
The initial and final energies and momenta are related by the usual conservation laws
\begin{equation}
E_1 + E_2 - E_1' - E_2' = 0 = p_1 + p_2 - p_1' - p_2' 
\end{equation}
which fixes the values for $p_1'$ and $p_2'$ uniquely.  

\subsection{One-loop}

Our analysis in this subsection follows \cite{Staudacher:2004tk}.
The reader may wish to consult this paper for further background.
We treat the dilatation operator as a Hamiltonian, and study the resulting problem of magnon scattering. 
Introduce the wave function $\psi (l_1,l_2,\cdots)$ as follows
\bea
O=\sum_{l_0,l_1,l_2,\cdots}\psi (l_1,l_2,\cdots)O_B((l_0)_0,(l_1,...)_1,(l_2,\cdots)_2,\cdots)\,.
\eea
The magnons at $l_1$ (magnon 1) and $l_2$ (magnon 2) meet at the edge 
of an annulus of radius squared $r_0^2 = {c_o\over N}$.
Magnon $1$ stretches to the edge of an annulus of radius squared $r_1^2 = {c_1\over N}$ 
and magnon $2$ to the edge of an annulus of radius squared $r_2^2 = {c_2 \over N}$.  
We do not assume any ordering of $c_0, c_1$ and $c_2$.
These two magnons are well separated from the remaining magnons (so we can consider them on their own)
and are both taken to be $Y$ impurities, which is general enough for our study.
The time independent Schr\"odinger equation following from the one loop dilatation operator is
\bea
E\psi (l_1,l_2)=\left(2 r_0^2+ r_1^2 +r_2^2\right)\psi(l_1,l_2)
-r_0 r_1 \left(\psi(l_1-1,l_2)+\psi(l_1+1,l_2)\right)\cr
-r_0 r_2 (\psi(l_1,l_2-1)+\psi(l_1,l_2+1))\label{separated}
\eea
The equation (\ref{separated}) is valid whenever the magnons are not adjacent in the string word, i.e. when $l_2>l_1+1$.
If the magnons are adjacent, we find
\bea
E\psi (l_1,l_1+1)=\left(r_1^2 +r_2^2\right)\psi(l_1,l_1+1)-   r_0 r_1\psi(l_1-1,l_1+1)
-r_0 r_2 \psi(l_1,l_1+2)\,.\label{adjacent}
\eea
Make the following Bethe ansatz for the wave function
\bea
   \psi(l_1,l_2)=e^{ip_1 l_1+ip_2 l_2}+R_{12}\, e^{ip_1'l_1+ip_2'l_2}\,.
\eea
It is straight forward to see that this ansatz obeys (\ref{separated}) as long as
\bea
   E=2r_0^2 + r_1^2 + r_2^2 -r_0 r_1(e^{ip_1}+e^{-ip_1})
-r_0 r_2 (e^{ip_2}+e^{-ip_2})\label{evalue}
\eea
and
\bea
r_0 r_1 (e^{ip_1}+e^{-ip_1})+r_0 r_2 \left( e^{ip_2}+e^{-ip_2}\right)=
r_0 r_1 (e^{ip_1'}+e^{-ip_1'})+ r_0 r_2 \left(e^{ip_2'}+e^{-ip_2'}\right)
\,.\label{relatemom}
\eea
Note that (\ref{evalue}) is indeed the correct one loop anomalous dimension and (\ref{relatemom}) can be obtained by
equating the  $O(\lambda)$ terms in the equation expressing the conservation of magnon energies.
From (\ref{adjacent}) we can solve for the reflection coefficient $R_{12}$.  
The result is
\bea
   R_{12}=-{2e^{ip_2}-\frac{r_1}{r_0} e^{ip_1+ip_2}-\frac{r_2}{r_0} \over 2e^{ip_2'}-\frac{r_1}{r_0} e^{ip_1'+ip_2'}-\frac{r_2}{r_0}  }  \label{1LoopA}
\eea
This is a rather simple result.  
Two easy checks are
\begin{itemize}
\item[1.] We see that $R_{12}R_{21}=1$.
\item[2.] If we set $r_1=r_2=r_0=1$ we recover the S-matrix of \cite{Staudacher:2004tk}. If we set $r_0=r_2=N$
and keep $r_1$ arbitrary, we recover the result of \cite{Koch:2015pga}.
\end{itemize}

To continue we need to go beyond the $su(2)$ sector by considering a state with a single $Y$ impurity and a single $X$
impurity. The operator with a $Y$ impurity at $l_1$ (this is now magnon 1) and an $X$ impurity at $l_2$ (this is now
magnon 2) is denoted $O_B((l_0)_0,(l_1,...)_1,(l_2,\cdots)_2,\cdots)_{YX}$ and the operator with an $X$ impurity at 
$l_1$ (magnon 1) and a $Y$ impurity at $l_2$ (magnon 2) is denoted $O_B((l_0)_0,(l_1,...)_1,(l_2,\cdots)_2,\cdots)_{XY}$.
Introduce the pair of wave functions 
\bea
O=\sum_{l_1,l_2,\cdots}\left[\psi_{YX} (l_1,l_2,\cdots)O_B((l_0)_0,(l_1,...)_1,(l_2,\cdots)_2,\cdots)_{YX}\right.\cr
\left.+\psi_{XY} (l_1,l_2,\cdots)O_B((l_0)_0,(l_1,...)_1,(l_2,\cdots)_2,\cdots)_{XY}\right]\,.
\eea
We find the time independent Schr\"odinger equation (\ref{separated}) for each wave function, 
when the impurities are not adjacent.
When the impurities are adjacent, we find
\bea
E\psi_{AB}(l_1,l_1+1)=\left(r_0^2 + r_1^2 + r_2^2 \right)\psi_{AB}(l_1,l_1+1)
-r_0 r_1 \psi_{AB}(l_1-1,l_1+1)\cr
-r_0^2\psi_{BA}(l_1,l_1+1)-r_0 r_2\psi_{AB}(l_1,l_1+2) \label{sep1}
\eea
where $A$ and $B$ could be either $X$ or $Y$.  
Make the following Bethe ansatz for the wave function
\bea
\psi_{YX}(l_1,l_2)=e^{ip_1 l_1+ip_2 l_2}+Ae^{ip_1'l_1+ip_2'l_2}\qquad
\psi_{XY}(l_1,l_2)=Be^{ip_1'l_1+ip_2'l_2}
\eea
The two equations of the form (\ref{separated}) imply that both $\psi_{XY}(l_1,l_2)$ and $\psi_{YX}(l_1,l_2)$
have the same energy, which is given in  (\ref{evalue}).
The equations (\ref{sep1}) imply that
\bea
A={e^{ip_2'}+e^{ip_2}-\frac{r_2}{r_0}-\frac{r_1}{r_0}e^{ip_2'+ip_1'}\over \frac{r_2}{r_0}+\frac{r_1}{r_0} e^{ip_1'+ip_2'}
-2e^{ip_2'}}\qquad
B={e^{ip_2'}-e^{ip_2}\over \frac{r_2}{r_0}+\frac{r_1}{r_0} e^{ip_1'+ip_2'}-2e^{ip_2'}}\,.
\eea
These again have a nice simple dependence on $c_0$, $c_1$ and $c_2$.
One can check that $|A|^2+|B|^2=1$ which is a consequence of unitarity.
To perform the check it is necessary to use the conservation of momentum $p_1+p_2=p_1'+p_2'$ and the
constraint (\ref{relatemom}).
We now obtain
\bea
   {A\over R_{12}}={e^{ip_2'}+e^{ip_2}-\frac{r_2}{r_0}-\frac{r_1}{r_0} e^{ip_1'+ip_2'}\over 
2e^{ip_1}-\frac{r_2}{r_0} -\frac{r_1}{r_0} e^{ip_1+ip_2}}\,.  \label{1LoopKin}
\eea
This should be equal to 
\bea
{1\over 2}\left( 1+{B_{12}\over A_{12}}\right)
\eea
where $A_{12}$ and $B_{12}$ are the exact S-matrix elements determined by the $su(2|2)^2$ symmetry.
We now expand the exact S-matrix in powers of $g$ and compare it with the above one-loop result. 
The first order expression is
\begin{eqnarray}
A_{12} & = & - \frac{2 e^{i p_2} - \frac{r_1}{r_0} e^{i(p_1 + p_2)} - \frac{r_2}{r_0} }{2 e^{i p_2'} - \frac{r_1}{r_0} e^{i(p_1' + p_2')} - \frac{r_2}{r_0} } \times \frac{\eta_1 \eta_2}{\eta_1' \eta_2'}  \label{ExactS1} \\
\frac{1}{2}\left( 1 + \frac{A_{12}}{B_{12}}  \right) & = & \frac{-e^{i p_2'} - e^{i p_2} + \frac{r_1}{r_0} e^{i(p_1 + p_2)} + \frac{r_2}{r_0}      }{ - 2 e^{i p_2} + \frac{r_1}{r_0} e^{i(p_1 + p_2)} + \frac{r_2}{r_0} } \label{ExactS2}.  
\end{eqnarray}
There is complete agreement between (\ref{1LoopKin}) and (\ref{ExactS2}).

\subsection{Two loops}

We now consider scattering to two loops.  
A similar computation for the $su(2)$ sector is considered in \cite{InelasticMagnons} to determine the energy and the
scattering element $R_{12}$ to two loops.  
At two loops $R_{12}$ receives non-trivial contributions from the overall scattering phase.  
To carry out a check independent of the unknown scattering phase, we again compare ratios of $S$-matrix elements. 
We can carry this out in the $su(2|3)$ sector of the theory.  
At two loops fermionic terms contribute.  
The various interaction terms in the dilatation operator up to two loops have been determined in \cite{su23}.  
For our study we begin by generalizing the existing result by adjusting the weights of the various terms.

Our initial state consists of two types of impuries so that only the following terms in the dilatation operator are 
relevant (we use the notation spelled out in \cite{su23})
\begin{eqnarray}
D_0 & = & \left\{ \begin{array}{c} a \\ a   \end{array} \right\} + \frac{3}{2}\left\{ \begin{array}{c} \alpha \\ \alpha   \end{array} \right\} 
\nonumber \\
D_2 & = & \alpha_1^2 \left\{ \begin{array}{cc} a & b \\ a & b  \end{array} \right\} 
+\alpha_1^2 \left(\left\{ \begin{array}{cc} a & \beta \\ a & \beta  \end{array} \right\}
+\left\{ \begin{array}{cc} \alpha & b \\ \alpha & b  \end{array} \right\}\right)
+\alpha_1^2 \left\{ \begin{array}{cc} \alpha & \beta \\ \alpha & \beta  \end{array} \right\}\cr
&&-\alpha_1^2 \left\{ \begin{array}{cc} a & b \\ b & a  \end{array} \right\} 
-\alpha_1^2 \left(\left\{ \begin{array}{cc} a & \beta \\ \beta & a \end{array} \right\}
+\left\{ \begin{array}{cc} \alpha & b \\  b & \alpha  \end{array} \right\}\right)
+\alpha_1^2 \left\{ \begin{array}{cc} \alpha & \beta \\ \beta & \alpha  \end{array} \right\}
\nonumber \\
D_3 & = & -\frac{1}{\sqrt{2}} \alpha_1^3 e^{i \beta_2} \epsilon_{\alpha \beta} \epsilon^{a b c} \left\{ \begin{array}{ccccc} & \alpha & & \beta & \\ a & & b & & c     \end{array} \right\}    -\frac{1}{\sqrt{2}} \alpha_1^3 e^{-i \beta_2} \epsilon^{\alpha \beta} \epsilon_{a b c} \left\{ \begin{array}{ccccc}  a & & b & & c \\ & \alpha & & \beta &    \end{array} \right\}  \nonumber \\
D_4 & = & (-2\alpha_1^4 + 2 \alpha_1 \alpha_3)\left( \left\{ \begin{array}{ccc} a & b & c \\ a & b & c \end{array}   \right\} \right) + (\frac{3}{2}\alpha_1^4 - \alpha_1 \alpha_3)\left( \left\{ \begin{array}{ccc} a & b & c \\ b & a & c \end{array}   \right\} + \left\{ \begin{array}{ccc} a & b & c \\ a & c & b \end{array}   \right\}  \right) \nonumber \\
& & - \frac{1}{2}\alpha^4 \left( \left\{ \begin{array}{ccc} a & b & c \\ b & c & a \end{array}   \right\} + \left\{ \begin{array}{ccc} a & b & c \\ c & a & b \end{array}   \right\}  \right)  \label{su23SC}
\end{eqnarray}
where $a, b$ and $c$ are any of the bosonic scalar fields and $\alpha$, $\beta$ are any of the fermionic fields of the
$su(2|3)$ sector. 
$D_0$ is the classical dilatation operator, $D_2$ a one loop contribution and $D_4$ a two loop contribution.
The term $D_3$ which is of order $g_{YM}^3$ mixes operators with a different number of fields since it allows
replacements of three boson fields with two fermion fields and vice versa.
Using the results developed in section $2$ we can adjust the weights of the various terms appearing in the dilatation
operator, based on the local factors for the magnons involved.
These weights are spelled out in our Bethe ansatz equations below.
We set $\alpha_1 = \frac{\sqrt{2} g}{\sqrt{N}}$ and $\alpha_3 = 0$.  
Factors of $N$ are absorbed by replacing $\sqrt{c_i} \rightarrow r_i$.

We use $\psi_{AB}$ for a wave function with $A$ and $B$ bosonic excitations 
($A$ and $B$ could be either $X$ or $Y$) and $\psi_{ab}$ for a wave function with fermionic excitations.  
When the magnons are well separated $l_2 > l_1 + 2$ we find

\begin{eqnarray}
& & E \psi_{AB}(l_1, l_2) \nonumber \\
& = & 2 g^2\left(2 r_0^2 + r_1^2 + r_2^2 \right)\psi_{AB}(l_1, l_2) \nonumber \\
& &  -  2 g^2 r_0 r_1 \left( \psi_{AB}(l_1 - 1, l_2) + \psi_{AB}(l_1 + 1, l_2) \right) - 2 g^2 r_0 r_2 \left( \psi_{AB}(l_1, l_2+1) + \psi_{AB}(l_1, l_2-1) \right) \nonumber \\
&  & - 2 g^4 \left( (r_0^2 + r_1^2)^2 + (r_0^2 + r_2^2)^2 \right) \psi_{AB}(l_1, l_2) + 4 g^4 \left( r_0^2 + r_1^2 \right) r_0 r_1 \left( \psi_{AB}(l_1+1, l_2) + \psi_{AB}(l_1-1, l_2) \right) \nonumber \\
& & + 4 g^4 \left( r_0^2 + r_2^2 \right) r_0 r_2 \left( \psi_{AB}(l_1, l_2+1) + \psi_{AB}(l_1, l_2-1) \right) \nonumber \\
& &  - 2 g^4 r_0^2 r_1^2\left( \psi_{AB}(l_1+2, l_2) + 2 \psi_{AB}(l_1, l_2) + \psi_{AB}(l_1-2, l_2) \right) \nonumber \\
& &  - 2 g^4 r_0^2 r_2^2\left( \psi_{AB}(l_1, l_2+2) + 2 \psi_{AB}(l_1, l_2) + \psi_{AB}(l_1, l_2-2) \right) + o(g^6) \label{l2Big}
\end{eqnarray}
When the magnons are seperated by two sites i.e. $l_2 = l_1+2$ we have
\begin{eqnarray}
& & E \psi_{AB}(l_1, l_1+2) \nonumber \\
& = &  2 g^2\left(2 r_0^2 + r_1^2 + r_2^2 \right)\psi_{AB}(l_1, l_1+2) \nonumber \\
& &  -  2 g^2 r_0 r_1 \left( \psi_{AB}(l_1 - 1, l_1+2) + \psi_{AB}(l_1 + 1, l_1+2) \right) - 2 g^2 r_0 r_2 \left( \psi_{AB}(l_1, l_1+3) + \psi_{AB}(l_1, l_1+1) \right) \nonumber \\
&  & - 2 g^4 \left( (r_0^2 + r_1^2)^2 + (r_0^2 + r_2^2)^2 \right) \psi_{AB}(l_1, l_1+2) \nonumber \\ 
& & + 4 g^4 \left( r_0^2 + r_1^2 \right) r_0 r_1 \left( \psi_{AB}(l_1+1, l_1+2) + \psi_{AB}(l_1-1, l_1+2) \right) \nonumber \\
& & + 4 g^4 \left( r_0^2 + r_2^2 \right) r_0 r_2 \left( \psi_{AB}(l_1, l_1+3) + \psi_{AB}(l_1, l_1+1) \right) \nonumber \\
& & -2 g^4 r_0^2 r_1^2\left( 2 \psi_{AB}(l_1, l_1+2) + \psi_{AB}(l_1-2, l_1+2) \right) \nonumber \\
& &  - 2 g^4 r_0^2 r_2^2\left( \psi_{AB}(l_1, l_1+4) + 2 \psi_{AB}(l_1, l_1+2) \right) - 4 r_0^4 \psi_{AB}(l_1, l_1+2)   \nonumber \\
& & + 2 g^4 r_0^2 \left( r_0 r_2 \psi_{AB}(l_1, l_1+1) + r_0 r_1 \psi_{AB}(l_1+1, l_1+2)  \right) \nonumber \\
& & - 2 g^4(r_0^3 r_2 \psi_{BA}(l_1, l_1 + 1) + r_0^3 r_1 \psi_{BA}(l_1+1, l_1+2)) \nonumber \\
& & -2 g^3 r_0^3 e^{-i\beta} (-\psi_{ab}(l_1, l_1 + 1) + \psi_{ba}(l_1, l_1 + 1)) + o(g^5) \label{l2Two}
\end{eqnarray}
and when they are adjacent i.e. $l_2 = l_1+1$
\begin{eqnarray}
& & E \psi_{AB}(l_1, l_1+1) \nonumber \\
& = & 2 g^2 \left(r_0^2 + r_1^2 + r_2^2 \right)\psi_{AB}(l_1, l_1+1) -  2 g^2 r_0 r_1 \left( \psi_{AB}(l_1 - 1, l_1+1) \right) - g^2 r_0 r_2 \left( \psi_{AB}(l_1, l_1+2) \right) \nonumber \\
&  & -2 r_0^2 g^2 \psi_{BA}(l_1, l_1 + 1) \nonumber \\
&  & - 2 g^4 \left( ( r_1^2 + r_0^2)^2 + (r_2^2 + r_0^2)^2 \right) \psi_{AB}(l_1, l_1+1) + 4 g^4 (r_0^2 + r_1^2 )r_0 r_1 \psi_{AB}(l_1-1, l_1+1) \nonumber \\
& & + 4 g^4 (r_0^2 + r_2^2)r_0 r_2  \psi_{AB}(l_1, l_1+2) - 2 g^4 r_0^2 r_1^2\left( \psi_{AB}(l_1, l_1+1) + \psi_{AB}(l_1-2, l_1+1) \right) \nonumber \\
& &  - 2 g^4 r_0^2 r_2^2\left( \psi_{AB}(l_1, l_1+1) + \psi_{AB}(l_1, l_1+3) \right) - 2 g^4 r_0^2 r_1 r_2 \left( \psi_{AB}(l_1-1, l_1) + \psi_{AB}(l_1+1, l_1+2) \right) \nonumber \\
& & - 4 g^4 \left( r_0^2 \psi_{BA}(l_1, l_1+1) \right) + 2 g^4 r_0^3 r_1 \psi_{AB}(l_1-1, l_1+1) + 2 g^4 r_0^3 r_2 \psi_{AB}(l_1, l_1+2)\nonumber \\
& &  - 2 g^4\left( r_0^3 r_1 \psi_{BA}(l_1 - 1, l_1+1) - 2(r_0^4 + r_0^2 r_1^2 + r_0^2 r_2^2) \psi_{BA}(l_1, l_1 + 1) + r_0^3 r_2 \psi_{BA}(l_1, l_1+2)  \right) \nonumber \\
& & -2 e^{-i\beta} g^3 r_0^2\left( r_1 \psi_{ab}(l_1 - 1, l_1) + r_2 \psi_{ab}(l_1, l_1+1) - r_1 \psi_{ba}(l_1-1, l_1) - r_2 \psi_{ba}(l_1, l_1+1)\right). \label{l2One} 
\end{eqnarray}
We also require the two-loop action on the fermionic excitations. 
When the fermionc excitations are adjacent we find

\begin{eqnarray}
& & E \psi_{ab}(l_1, l_1+1) \nonumber \\
& = &  2 g^2  (r_0^2 + r_1^2 + r_2^2)\psi_{ab}(l_1, l_1+1) - 2 g^2 r_0 r_1 \psi_{a b}(l_1-1, l_1+1)  - 2 g^2 r_0 r_2 \psi_{a b}(l_1, l_1+2)   \nonumber \\
& &  + 2 g^2 r_0^2 \psi_{ba}(l_1, l_1+1) \nonumber \\
& &  - 2 g^3e^{i \beta} r_0^2 \left( r_2 \psi_{AB}(l_1, l_1+1) - r_2 \psi_{BA}(l_1, l_1+1) - r_0 \psi_{AB}(l_1, l_1+2) \right. \nonumber \\
& & + r_0 \psi_{BA}(l_1, l_1+2) + r_1 \psi_{AB}(l_1+1, l_1+2) - r_1 \psi_{BA}(l_1+1)  + o(g^5). \label{Ferml2} 
\end{eqnarray}
To solve the above equations make the following Bethe ansatz \cite{InelasticMagnons,ReflectingMagnons}
\begin{eqnarray}
\psi_{YX}(l_1, l_2) & = & e^{i p_1 l_1 + i p_2 l_2} + A e^{i p_3 l_1 + i p_4 l_2} + r_0^2 g^2 \delta_{l_2, l_1+1}  \phi_{YX}(l_1) \nonumber \\
\psi_{XY}(l_1, l_2) & = & B e^{i p_3 l_1 + i p_4 l_2} + r_0^2 g^2 \delta_{l_2, l_1+1} \phi_{XY}(l_1) \nonumber \\
\psi_{a b}(l_1, l_2) & = & g C_1 e^{i p_3 l_1 + i p_4 l_2} \qquad
\psi_{b a}(l_1, l_2)  =  g C_2 e^{i p_3 l_1 + i p_4 l_2}.  
\end{eqnarray}
Since we are working to two loops we can solve for $p_1'$ and $p_2'$ perturbatively to find
\begin{eqnarray}
p_1' & = & -i \log\left( \frac{e^{ip_2}(r_1 + e^{ip_1 + i p_2}r_2) }{e^{ip_1 + ip_2}r_1 + r_2} +  \right) +  4 g^2 r_1 r_2 (r_1^2 - r_2^2) \times \nonumber \\
& &  \times e^{2 i (p_1 + p_2)} \left( \frac{r_1^2 + r_2^2 - 2 r_0(r_1 \cos(p_1) + r_2 \cos(p_2) ) + 2 r_1 r_2 \cos(p_1 + p_2)} {(e^{i(p_1 + p_2)}r_1 + r_2)^2 (e^{i(p_1 + p_2)}r_2 + r_1)^2   } \right)\sin(p_1 + p_2) \nonumber 
\end{eqnarray} 
and $p_2' = p_1 + p_2 - p_1'$.
Using these expressions and plugging our ansatz into the scattering equations we find
\begin{eqnarray}
E & = & 2(r_0^2 + r_1^2 - 2 r_0 r_1 \cos(p_1))g^2 - 2 (r_0^2 + r_1^2 - 2 r_0 r_1 \cos(p_1))^2 g^4  \nonumber \\ 
& & + 2(r_0^2 + r_2^2 - 2 r_0 r_2 \cos(p_2))g^2 - 2 (r_0^2 + r_2^2 - 2 r_0 r_2 \cos(p_2))^2 g^4 + o(g^5).  
\end{eqnarray}
which solves (\ref{l2Big}).  
This energy agrees with the exact expression to two loops.
From (\ref{Ferml2}) we determine $C_1$ and $C_2$ and  from (\ref{l2Two}) we obtain 
$\phi_{XY}$ and $\phi_{YX}$ in terms of $A$ and $B$.  
These expressions, given in Appendix \ref{klunky}, are cumbersome and are not quoted.
Finally (\ref{l2One}) determines $A$ and $B$.  
The results are in Appendix \ref{klunky}.  
In the end we obtain
\begin{eqnarray}
		A & = & \frac{e^{i p_2} r_0 r_1 + e^{i (2 p_1 + p_2)} r_0 r_1 + e^{i p_1} r_0 r_2 + 
 e^{i (p_1 + 2 p_2)} r_0 r_2 - r_1 r_2 - e^{2 i (p_1 + p_2)} r_1 r_2 - 
 e^{i (p_1 + p_2)} (r_1^2 + r_2^2)}{(e^{i (p_1 + p_2)} r_1 + 
   r_2) (-2 e^{i p_1} r_0 + r_1 + e^{i (p_1 + p_2)} r_2)} \nonumber\\
	& & + 2 e^{-i (p_1 + p_2))} r_0 \frac{N_A}{D_A}g^2	+ o(g^4)   \label{A2Loop}
		\end{eqnarray}
and
\begin{equation}
B = \frac{r_0 (e^{i p_2} r_1 - e^{i (2 p_1 + p_2)} r_1 - e^{i p_1} r_2 + 
   e^{i (p_1 + 2 p_2)} r_2}{(e^{i (p_1 + p_2)} r_1 + r_2) (-2 e^{i p_1} r_0 +
    r_1 + e^{i (p_1 + p_2)} r_2)} -2 e^{-i p_2} r_0 \frac{N_B}{D_A} g^2 + o(g^4)   \label{B2Loop}
\end{equation}
When $r_0 = r_1 = r_2 = 1$ we recover the scattering matrix of \cite{Beisert} and when $r_0 = r_1 = 1$ 
we recover the scattering matrix of \cite{InelasticMagnons}.
We now find
\begin{equation}
\frac{A}{A + B} = \frac{1}{2}\left(1 + \frac{B_{12}}{A_{12}} \right) 
\end{equation}
for any choice of $r_1, r_2$ and $r_0$.  
This demonstrates that our $S$-matrix agrees with the CFT scattering to two loops. 

When $r_1 = r_2 = r_0$ we find the simpler expressions
{\footnotesize
\begin{eqnarray}
A & = & -\frac{(-1 + e^{i p_1}) (-1 + e^{i p_2})}{
  1 - 2 e^{i p_1} + e^{i (p_1 + p_2)}} \nonumber \\
	& & + \frac{
 2 e^{-i (p_1 + p_2)} (-1 + e^{i p_1})^2 (-1 + e^{i p_2})^2 (-e^{i p_1} + 
    e^{2 i p_1} + e^{i p_2} - e^{i (p_1 + p_2)} - e^{i (2 p_1 + p_2)} + e^{
    i (p_1 + 2 p_2)}) r_0^2 g^2}{(1 - 2 e^{i p_1} + e^{i (p_1 + p_2)}   )^2} \nonumber \\
B & = & \frac{-e^{i p_1} + e^{i p_2}}{1 - 2 e^{i p_1} + e^{i (p_1 + p_2)}} + \frac{
 2 e^{-i p_2} (-1 + e^{i p_1})^2 (-1 + e^{i p_2})^2 (-e^{i p_1} + e^{
    i p_2}) r_0^2 g^2}{(1 - 2 e^{i p_1} + e^{i (p_1 + p_2)})^2}.  		
\end{eqnarray}}
These S-matrices are related to those of \cite{Beisert} by rescaling $g^2 \rightarrow g^2 r_0^2$, as
predicted in \cite{Koch:2016jnm}.  

\section{Finite Size Corrections}

We now develop the string theory description of the gauge theory magnons.
This probes the strong coupling limit of the magnon dynamics. 
We will compute corrections to the energy of magnons that carry finite angular momentum.
As explained in the introduction, these finite size corrections are not determined by the $su(2|2)^2$
symmetry of the problem.  

We study classical string solutions on the $\left\{ t,r, \phi \right\}$ subspace of an LLM background
\bea
ds^2 & = & \left( r^2 h^2(r) - h^{-2}(r) (1 + V_{\phi}(r))^2  \right) dt^2 + 2(r^2 h^2(r) - 2 V_{\phi} h^{-2}(r)(1 + V_{\phi}(r))) d\phi dt     \nonumber \\
     &   &   + \left( r^2 h^2(r) - h^{-2}(r) V_{\phi}(r)^2  \right) d\phi^2 + h^2(r) dr^2.  
\eea
We will work with the Nambu-Goto action
\bea
S_{NG} = \frac{\sqrt{\lambda}}{2\pi} \int \sqrt{(\dot{X} \cdot X')^2 - \dot{X}\cdot\dot X  X' \cdot X'}.  
\eea
The giant magnon excitation of a string in the $AdS_5 \times S^5$ background can be found as a solution to the 
classical equations of motion\cite{GiantMagnons}.  
Remarkably, the giant magnon
\bea
r(\tau, \sigma) = c \sec\left( \phi(\tau, \sigma) + t(\tau, \sigma) + \phi_0 \right)
\eea
also solves the equations of motion for any LLM geometry.  
Here $c, \phi_0$ are integration constants fixed by appropriate boundary conditions\footnote{The rescaling predicted 
by \cite{Koch:2016jnm} naturally appears in these boundary conditions.  This will be explicit in the form of the 
finite size corrections}.  
It is convenient to choose the worldsheet coordinates as $\phi = \sigma - \tau$ and $t = \tau$ so that $r = r(\sigma)$.  
The giant magnon maintains its shape and is rotating at an angular velocity of $1$.

To compute the finite size corrections we need to construct a finite angular momentum magnon solution.
The finite magnon \cite{FiniteSize} is a classical string solution that maintains its shape but is rotating at an angular 
velocity less than $1$.  
Our ansatz for the finite magnon is
\bea
r = r(\kappa \phi(\tau, \sigma) + t(\tau, \sigma)  ).  
\eea
A convenient choice of worldsheet coordinates is $\phi = \sigma - \tau$ and $t = \kappa \tau$ so that again $r = r(\sigma)$.
The equations of motion can be integrated to find
\bea
r'(\sigma) = \frac{\kappa r \sqrt{1 - \frac{r^2}{C^2}}}{\sqrt{(1 - \kappa)^2 h^{4}(r) r^2 - (\kappa - (1 - \kappa)V_{\phi}(r))^2}}  \label{rpEq}
\eea
where $C$ is an integration constant.  Note that when $\kappa = 1$ we recover the giant magnon solution.
The conserved charges are computed as 
\bea
E = \frac{\sqrt{\lambda}}{2\pi}\int_{\sigma_{\min}}^{\sigma_{max}} d\sigma \frac{\partial L_{NG}}{\partial \dot{t}}  
\qquad
J = \frac{\sqrt{\lambda}}{2\pi}\int_{\sigma_{\min}}^{\sigma_{max}} d\sigma \frac{\partial L_{NG}}{\partial \dot{\phi}}, 
\label{IsToDo} 
\eea
with the right hand sides of these expressions evaluated on the classical solution.
To evaluate these integrals it is useful to make the coordinate transformation from $\sigma$ to $r$ using (\ref{rpEq}).
As an example, the expression for the conserved momentum simplifies significantly
\begin{eqnarray}
p(\kappa, C) & = & \int_{\sigma_{min}}^{\sigma_{max}} d\sigma =  p_1 + p_2 \nonumber \\
\textnormal{where} \ \ \ \ \ p_i & = & \left|\int_{C}^{r_{max_i}} \frac{1}{r'(\sigma)} dr \right| \label{pInt}
\end{eqnarray}
where $r_{max_i}$ are the endpoints of the string.  
These are any two roots of the denominator of (\ref{rpEq}) and are thus functions of $\kappa$.  
It is useful to define the integral
\begin{equation}
R_{p_i} = \left|\int_{C}^{r_{max_i}} \frac{r^2}{r'(\sigma)} dr \right| 
\end{equation}
The evaluation of (\ref{IsToDo}) requires further changes of integration variables.
Consider the case $r_{max_i} \approx r_i$ so that when $\kappa = 1$ the magnon's endpoint is at $r_i$.
In this case, the relevant change of variables is
\begin{eqnarray}
r & = & r_i \sqrt{1 - \tilde{C_i}^2 z^2} \qquad
C  =  r_i \sqrt{1 - \tilde{C_i}^2} \nonumber \cr\cr
\kappa(f) & = & \frac{1}{1 + \left[(r_i \sqrt{1 - \tilde{C_i}^2 f^2}) h^{2}(z = f) + V_{\phi}(z = f) \right]^{-1}}.  
\end{eqnarray}
The $z$-integral now runs from $f$ to $1$ for any LLM geometry, allowing a general evaluation.  
We find the following relations between the conserved charges
\begin{eqnarray}
E_i & = & \frac{1}{\kappa(f)} J_i + \frac{\sqrt{\lambda}}{2 \pi}\frac{R_{p_i}}{ 2 C } \label{ECharge}  \\
J_i & = & \frac{\sqrt{\lambda}}{2 \pi} \frac{1}{2 C} \frac{\kappa^2}{\partial_f \kappa} \partial_f 
\left( R_{p_i} - C^2 p_i \right),  \label{JCharge}
\end{eqnarray}
which can be verified at the level of the integrands.  
The derivative with respect to $f$ can be taken outside of the integrals since $\frac{1}{z'(\sigma)}$ vanishes at $z = f$.  
We thus only need the integrals $p_i$ and $R_{p_i}$ to evaluate the conserved charges.
For our problem of concentric rings, the edge of the ring is at $r = r_i$, where the function $V_\phi(r)$ has a pole.
Consequently, we can make the following substitution
\begin{eqnarray}
V_{\phi}(z) & = & \frac{1}{\tilde{C_i}^2 z^2}\left( 1 - \tilde{C_i}^2 z^2 + \bar{V}_{\phi, i}(z) \right) \qquad
h(z)  =  \frac{1}{\tilde{C} z}\left( 1 + \bar{h_i}(z)\right) \nonumber \\
\bar{V}_{\phi,i}(0) &=& \bar{h_i}(0) = 0
\end{eqnarray} 

In summary, we have two integrals to evaluate, $p_i$ and $R_{p_i}$.  
These integrals run from $f$ to $1$ with an integrand also depending on $f$.  
In Appendix \ref{evaluate} we develop a systematic expansion of these integrals in powers of $f$.

\subsection{Magnons attached to a single corner}

In this case $r_{max_1} = r_{max_2} \equiv r_{max} \approx r_0$, so the $\kappa=1$ magnon 
starts and ends at $r = r_0$.  
Collecting the results from the Appendix \ref{evaluate} we find, to second order in $f$, that
\begin{eqnarray}
\frac{p}{2} = p_1 = p_2 & = & \frac{\pi}{2} + i\log(\tilde{C} + i \sqrt{1 - \tilde{C}^2}) - \frac{1}{4}\left(  \tilde{C} \sqrt{1 - \tilde{C}^2}\left( 1 + 2 \log\left( \frac{4}{f} \right) \right)  \right)f^2 + I_4 f^2  \nonumber \\
R_{p_1} = R_{p_2} & = & \tilde{C}\sqrt{1 - \tilde{C}^2} r_0^2 - \frac{1}{4}\tilde{C}\sqrt{1 - \tilde{C}^2 }r_0^2\left( 1 - 2 \tilde{C}^2 + 2 \log\left( \frac{4}{f} \right) \right)f^2 + f^2 I_5  \label{pFin}
\end{eqnarray}
The finite size effects are evaluated by inserting these expressions for $p$ and $R_p$ into (\ref{ECharge}) and (\ref{JCharge}).  
The result, to second order in $f$, is
\begin{equation}
E - J = \frac{\sqrt{\lambda}}{\pi} r_0 \sin\left(\frac{p}{2} \right) - \frac{\sqrt{\lambda}}{4\pi} r_0 \tilde{C}^3 f^2 + o(f^4)
\end{equation}
This result is independent of the specific LLM geometry and is sensitive only to the radius $r_0$.
All that remains is the determination of $\tilde{C}$ and $f$ in terms of the leading order central charges, 
$p_0, E_0$ and $J_0$.  From the leading order in (\ref{pFin}) we find
\begin{equation}
\tilde{C} = \sin\left(\frac{p_0}{2} \right)
\end{equation}
To find $f$ we consider $E$ or $J$.  
Using the expression for $E$ to leading order in $f$ we find
\begin{eqnarray}
\frac{\pi}{\sqrt{\lambda}} E_0 & = & -r_0 \tilde{C} \log\frac{f}{4} + \int_{C}^{r_0} dr \frac{\sqrt{r^2 - C^2}}{r}\left[ V_{\phi}(r) - \frac{r^2}{r_0^2 - r^2} \right] \nonumber \\
& \equiv & -r_0 \tilde{C} \log\frac{f}{4} + \frac{\pi}{\sqrt{\lambda}} V,
\end{eqnarray}
which implies
\begin{equation}
f = 4 e^{-\frac{\pi(E_0 - V)}{\sqrt{\lambda}r_0 \sin(\frac{p}{2})}  } = 4 e^{-\left(\frac{\pi(J_0 - V)}{\sqrt{\lambda}r_0 \sin(\frac{p}{2})} + 1\right)  }
\end{equation}
The final result for the finite size effects is
\begin{equation}
E - J = \frac{\sqrt{\lambda}}{\pi} r_0 \sin\left(\frac{p}{2} \right) - 4\frac{\sqrt{\lambda}}{\pi} r_0 \sin^3(\frac{p}{2}) e^{-2\left(\frac{\pi(J_0 - V)}{\sqrt{\lambda}r_0 \sin(\frac{p}{2})} + 1\right)  } + o(f^4)  \label{FiniteSize}
\end{equation}
To interpret this answer, evaluate $J_0 - V$. First consider the $AdS_5 \times S_5$ example and set $\kappa = 1$ 
i.e. $f = 0$.  In this case $J_0$ is given by the (divergent) integral
\begin{equation}
J_0 \rightarrow \frac{\sqrt{\lambda}}{\pi}\int_{C}^{1}  \frac{\sqrt{r^2 - C^2}}{r} \left( \frac{r^2}{1 - r^2} \right) \ dr.  \label{AdS5J0}
\end{equation}
The above integral determines the length of the operator when we send the length to infinity.
For a general $LLM$ geometry at $\kappa = 1$ we find
\begin{equation}
J_0 \rightarrow \frac{\sqrt{\lambda}}{\pi}\int_{C}^{r_0}  \frac{\sqrt{r^2 - C^2}}{r}  V_{\phi}(r) \ dr. 
\end{equation}
Thus, due to the presence of $V_{\phi}(r)$ in the integrand, $J_0$ receives contributions from every ring of the 
LLM geometry.  After subtracting the factor $V$, we find 
\begin{equation}
J_0 - V \rightarrow \frac{\sqrt{\lambda}}{\pi}\int_{C}^{r_0}  \frac{\sqrt{r^2 - C^2}}{r} \left( \frac{r^2}{r_0^2 - r^2} \right) \ dr
\end{equation}
which only receives a contribution from the local geometry of the magnon.
This is expected for a quantity associated to the {\it local} magnon excitation.
Thus, the answer obtained (\ref{AdS5J0}) is sensitive only to the local geometry the string experiences.
Further, $J_0-V$ computes the number of fields in the dual CFT operator.
It is clear from (\ref{FiniteSize}) that both the dispersion relation for the giant magnon and its finite size effects 
agree with the AdS$_5\times$S$^5$ result after rescaling $\sqrt{\lambda}\to r_0\sqrt{\lambda}$.  
This is in precise agreement with the prediction of \cite{Koch:2016jnm}.

\subsection{Magnons stretching between different corners}

We now investigate the finite size effects for a magnon stretching between different corners.  
The relevant integrals are
\begin{eqnarray}
p_i & = & \left(\frac{\pi}{2} + i \log\left( i \frac{C}{r_i} + \sqrt{1 - \frac{C^2}{r_i^2}} \right) \right) - \frac{C}{4 r_i}\sqrt{1 - \frac{C^2}{r_i^2}}(1 + 2 \log(4) - 2 \log(f))f^2 + f^2 I_{4,i} \nonumber \\
R_{p_i} &=  & r_i C\sqrt{1 - \frac{C^2}{r_i^2}} - \frac{C\sqrt{1 - \frac{C}{r_i^2}}}{4 r_i}\left( -r_i^2 + 2 C^2 + 2 r_i^2 \log(4) - 2  r_i^2 \log(f^2) \right)f^2 + I_{5,i} f^2.    
\end{eqnarray}
To leading order in $f$ we find
\begin{equation}
C = \frac{r_1 r_2 \sin(p_0)}{r_1^2 + r_2^2 - 2 r_1 r_2 \cos(p_0)}.  
\end{equation}
To leading order in $E-J$ we obtain
\begin{eqnarray}
E_0 - J_0 & = & \frac{\sqrt{\lambda}}{2\pi}\left( \sqrt{r_1^2 - C^2} + \sqrt{r_2^2 - C^2} \right) \nonumber \\
& = & \frac{\sqrt{\lambda}}{2 \pi} \sqrt{r_1^2 + r_2^2 - 2 r_1 r_2 \cos(p_0)}. 
\end{eqnarray}
This reproduces the large coupling limit of the expression for the energy of a magnon stretched between different corners,
given in (\ref{Dim2Corners}).  
The finite size effect is given by
\begin{eqnarray}
& & E - J \nonumber \\
 & = & \frac{\sqrt{\lambda}}{2 \pi} \sqrt{r_1^2 + r_2^2 - 2 r_1 r_2 \cos(p)} \nonumber \\& &  + \frac{\sqrt{\lambda}}{2\pi} \frac{4(r_1^2 + 3 r_1^2 r_2^2 + r_2^4) + r_1 r_2(-15(r_1^2 + r_2^2)\cos(p) + 12 r_1 r_2 \cos(2 p) - (r_1^2 + r_2^2) \cos(3p) )}{16(r_1^2 + r_2^2 - 2 r_1 r_2 \cos(p))^{\frac{3}{2}}}f^2 \nonumber.  
\end{eqnarray}
The value for $f$ can be extracted from the leading order expression for $J$ given by 
\begin{equation}
J_0 - V_1 - V_2 = -\frac{\sqrt{\lambda}}{2\pi} \sqrt{r_1^2 + r_2^2 - 2 r_1 r_2 \cos(p)}\left( 1 + \log\left( \frac{f}{4} \right)\right).  
\end{equation}
Setting $r_1 = r_2 = r_0$ we recover (\ref{FiniteSize}).  
Notice that the answer depends on $J_0 - V_1 - V_2$.  
This is again as it must be for a local quantity: $J_0$ receives a contribution from all corners while $J_0 - V_1 - V_2$ 
receives contributions only from the corners where the magnon starts and ends.    

\subsection{Dyonic Giant Magnons}

The giant magnon solutions considered above are dual to single trace operators with $o(1)$ impurities.  
We now consider operators with $J$ fields, of which $J_1$ are $Y$-impurities, with $J_1$ scaling as $o(\sqrt{N})$.
These are dual to strings which have two conserved angular momentum so that an additional angle participates
when solving the equations of motion.  
We thus consider the $t, r, \phi, \theta$ subspace on the LLM plane
\begin{eqnarray}
ds^2 & = & \left( r^2 h^2(r) - h^{-2}(r) (1 + V_{\phi}(r))^2  \right) dt^2 + 2(r^2 h^2(r) - 2 V_{\phi} h^{-2}(r)(1 + V_{\phi}(r))) d\phi dt     \nonumber \\
     &   &   + \left( r^2 h^2(r) - h^{-2}(r) V_{\phi}(r)^2  \right) d\phi^2 + h^2(r) dr^2 + h^{-2}(r) d\theta^2.  
\end{eqnarray}
We find the following solution of the Nambu-Goto equations of motion
\begin{eqnarray}
\phi & = & \cos^{-1}\left( \frac{c}{r(\sigma, \tau)}\right) - t(\sigma, \tau) \nonumber \\
\theta & = & a\left( t(\sigma, \tau) + c F(r(\sigma, \tau))   \right)
\end{eqnarray}
where 
\begin{equation}
F'(r) = \frac{V_{\phi}(r)}{r\sqrt{r^2 - c^2}}.  
\end{equation}
A convenient choice of worldsheet coordinates is $t = \tau$ and $r = c \sec(\sigma)$.  
Change integration variables from $\sigma$ to $r$ and integrate from $r_m$ to $r_i$ and $r_m$ to $r_j$.

By inverting the definition for $p$ we find that
\begin{equation}
r_m = \frac{r_i r_j \sin(p)}{\sqrt{r_i^2 + r_j^2 - 2 r_i r_j \cos(p)}}
\end{equation}
In addition to the conserved charges $E$ and $J$, there is another conserved charge
\begin{equation}
J_2 \equiv \frac{\lambda}{2\pi} \int_{\sigma_{min}}^{\sigma_{max}} \frac{\partial L_{NG}}{\partial \dot{\theta}}.  
\end{equation}
Using these definitions and the solution given above we readily find
\begin{eqnarray}
E - J & = & \frac{\sqrt{\lambda}}{2 \pi \sqrt{1 - a^2} } \left( \sqrt{r_i^2 - r_m^2}   + \sqrt{r_j^2 - r_m^2}    \right) \nonumber \\
J_2 & = & -\frac{a \sqrt{\lambda}}{2 \pi \sqrt{1 - a^2} }\left( \sqrt{r_i^2 - r_m^2}   + \sqrt{r_j^2 - r_m^2}    \right)    .  
\end{eqnarray}
Solving for $a$ in terms of $J_2$ then yields the correct dispersion relation
\begin{equation}
E - J = \sqrt{J_2^2 + \frac{\lambda }{4 \pi^2}(r_i^2 + r_j^2) - \frac{\lambda}{2\pi^2} r_i r_j \cos(p)   }
\end{equation}
Notice that when $r_i = r_j$ the above dispersion relations are related to those for excitations in the AdS$_5\times$S$^5$
background by a simple rescaling of the coupling.  

\subsection{Finite size Dyonic Giant Magnon}

To go from the infinite size dyonic giant magnon to the finite size version we agin need to adjust the angular velocity of the string endpoints.  Make the ansatz
\begin{eqnarray}
t & = & \kappa \tau \qquad \phi  = \sigma - \tau \nonumber \\
r & = & r(\sigma) \qquad \theta = a(\tau + f(r(\sigma)))
\end{eqnarray}
After some work we find that the equations of motion are solved by
{\footnotesize
\begin{eqnarray}
f'(r(\sigma)) & = & \frac{-r^2(-1 + \kappa)h^{4}(r) + V_{\phi}(r)(\kappa + (-1 + \kappa)V_{\phi}(r) )}{r'(\sigma)\left(-r^2(-1 + \kappa)h^{4}(r) + (\kappa + (-1 + \kappa)V_{\phi}(r) )^2 \right)} \nonumber \\
r'(\sigma) & = & \frac{r \sqrt{-\kappa^4 + r^2(a - \kappa^2)C_1 + (-1 + \kappa)(\kappa^2 + r^2 C_1)\left( r^2(-1 + \kappa)h^4(r) + V_{\phi}(r)(-2\kappa + V_{\phi}(r) - \kappa V_{\phi}(r))   \right) }}{(\kappa + r(-1 + \kappa)h^2(r) + (-1 + \kappa)V_{\phi}(r)    )(-\kappa + r(-1 + \kappa)h^2(r) + (1 - \kappa)V_{\phi}(r)    )}.  \nonumber 
\end{eqnarray}}
When $\kappa = 1$ we recover the infinite size solution and when $a= 0$ we recover the finite size giant magnon solution.
Recall that for the finite size giant magnon we need the points where the denominator and numerator of $r'(\sigma)$ 
vanishes.  
Here we choose the constants $\kappa$ and $C_1$ so that the zeroes are at convenient values of $r$.  
The choice
\begin{eqnarray}
\kappa & = & 1 - \frac{1}{1 + r_{max} h^2(r_{max}) + V_{\phi}(r_{max})} \nonumber \\
C_1 & = & \frac{\kappa^2(-r_{min}^2(1 - \kappa)^2 h^{4}(r_{min}) + (\kappa - (1- \kappa) V_{\phi}(r_{min}) )^2 )}{r_{min}^2(  r_{min}^2(1 - \kappa)^2 h^4(r_{min}) +  (a + \kappa - (1 - \kappa) V_{\phi}(r_{min}) )(a - \kappa + (1 - \kappa) V_{\phi}(r_{min}) )  )}\nonumber
\end{eqnarray}
yields a zero in the numerator at $r = r_{min}$ and a zero in the denominator at $r = r_{max}$.  
There is an additional zero in the numerator at $r \approx r_{max}$ so that $r'(\sigma)$ becomes imaginary as we 
approach $r_{max}$.  
Our integration limits are thus the two zeroes in the numerator.  
We will compute the location of the second zero explicitly in a small $f$ expansion.  
In what follows, make use of the identity
\begin{equation}
E - \frac{J}{\kappa} = \frac{\sqrt{-C_1}}{\kappa} R_p
\end{equation}
true for any of the LLM geometries we consider.  
Make the following coordinate changes and ansatz
\begin{eqnarray}
V_{\phi} & = & \frac{r^2}{r_0^2 - r^2}\left( 1 + \tilde{V}_{\phi}(r)  \right) \qquad
h = \frac{\sqrt{r_i}}{\sqrt{r_i - r^2}}\left( 1 + \tilde{h}(r)  \right) \nonumber \\
r & = & r_0\sqrt{1 - \tilde{C} z^2} \qquad
r_{min} = r_0\sqrt{1 - \tilde{C}} \qquad
r_{max}  =  r_0\sqrt{1 - \tilde{C} f^2}
\end{eqnarray}
so that the zeros in the numerator and denominator respectively are at $z = 1$ and $z = f$. 
The additional zero is at $z = f\sqrt{1 + \frac{a^2}{(1-a^2)\tilde{C}^2}} + c(f) f^2 \equiv \tilde{f} $.  
The limits of integration are thus $z = \tilde{f}$ and $z = 1$.  
The value of $c(f)$ is unimportant in the first few orders of a small $f$ expansion.  
Using the method outlined in the Appendix \ref{evaluate}, we find
{\footnotesize
\begin{eqnarray}
\frac{p}{2} & = & \sin^{-1}(\tilde{C}) - \frac{\sqrt{1 - \tilde{C}^2}}{4(1 - a^2) \tilde{C}}\left(  \tilde{C}^2 + a^2(1 - \tilde{C}^2)   + (a^2 - (1- a^2)\tilde{C}^2) \log\left( \frac{(a^2 + \tilde{C}^2 - a^2 \tilde{C}^2)f^2 }{ 16(1 - a^2)\tilde{C}^2     } \right)   \right) f^2  \nonumber \\
  &   & - f^2 \int_0^{1} dz  \frac{\sqrt{1 - \tilde{C}^2}(-a^2 z^2(1 - \tilde{C}^2 z^2)V_{\phi}(1) + (\tilde{C}^2(-1 + z^2) +  a^2(1 + \tilde{C}^2 - 2 \tilde{C}^2 z^2)   )V_{\phi}(z)    )    }{2 (1 - z^2)^{\frac{3}{2}}(1 - a^2)\tilde{C} z (-1 + \tilde{C}^2 z^2)     } \nonumber \\
\frac{R_p}{2} & = & \tilde{C}\sqrt{1 - \tilde{C}^2} r_i^2 \nonumber \\
& &  - \frac{\sqrt{1 - \tilde{C}^2} r_i^2}{4 (1 - a^2) \tilde{C}}\left( \tilde{C}^2 - 2\tilde{C}^4 + a^2(1 + \tilde{C}^2 + 2 \tilde{C}^4) + (a^2 - (1 - a^2)\tilde{C}^2) \log\left( \frac{(a^2 + \tilde{C}^2 - a^2 \tilde{C^2})(f^2)}{16(1-a^2) \tilde{C}^2} \right)   \right)   \nonumber \\
 & & + f^2 r_i^2 \int_0^{1} dz  \frac{\sqrt{1 - \tilde{C}^2}(-a^2 z^2(1 - \tilde{C}^2 z^2)V_{\phi}(1) + (\tilde{C}^2(-1 + z^2) +  a^2(1 + \tilde{C}^2 - 2 \tilde{C}^2 z^2)   )V_{\phi}(z)    )    }{2 (1 - z^2)^{\frac{3}{2}}(1 - a^2)\tilde{C} z     } \nonumber \\
\frac{\pi J_2}{\sqrt{\lambda}} & = & -\frac{a \tilde{C} r_i}{\sqrt{1 - a^2}} + \frac{a(-\tilde{C}^2 - a^2(1 - \tilde{C}^2)) r_i }{4(1 - a^2)^{\frac{3}{2}} \tilde{C}  }\left( -1 + \log\left(\frac{16(1 - a^2) \tilde{C}^2}{(\tilde{C}^2 + a^2(1 - \tilde{C}^2))f^2} \right) \right) \nonumber \\
 &  & - a r_i f^2 \int d z  \frac{a^2(-1 + \tilde{C}^2) z^2 V_{\phi}(1) + (a^2(1 - \tilde{C}^2) +  \tilde{C}^2(1 - z^2)   )V_{\phi}(z)   }{ 2 \tilde{C} z (1 - a^2)^{\frac{3}{2}} (1 - z^2)^{\frac{3}{2}}  } \nonumber \\
\frac{\pi J}{\sqrt{\lambda}} & = & \frac{r_i}{\sqrt{1 - a^2}\tilde{C}}\left( - \tilde{C}^2 + (\tilde{C}^2 + a^2(1 - \tilde{C}^2))\log\left( \frac{4 \tilde{C}\sqrt{1 - a^2}  }{\sqrt{\tilde{C}^2 + a^2(1 - \tilde{C}^2)} f}   \right) \right)  \nonumber \\
& & + \int_0^1 dz \frac{r_i\sqrt{1 - z^2}(a^2(1 - \tilde{C}^2 + \tilde{C}^2(1 - z^2)  ) V_{\phi}(z)}{ \tilde{C} z (1 - z^2)(1 - \tilde{C}^2 z^2)    }\nonumber
\end{eqnarray}}
The terms which depend on the details of the LLM geometry considered are all finite, which can be verified by performing an integration by parts.  
We find
\begin{equation}
E - J - \sqrt{J_2^2 + r_i^2 \frac{\lambda}{\pi^2} \sin^2( \frac{p}{2} )} = -\frac{\sqrt{\lambda}}{\pi}\frac{r_i}{4}\frac{\tilde{C}(\tilde{C}^2 + a^2(1 - \tilde{C}^2))}{\sqrt{1 - a^2}}f^2 + o(f^3)
\end{equation}
where terms sensitive to the details of the LLM geometry have canceled.  
Indeed, we have
\begin{equation}
- \int_0^1 dz \frac{a^2 \tilde r_i (1 - z^2)(z - \sqrt{1 - z^2})V_{\phi}(1) f^2   }{2( (1 - a^2)(1 - z^2)  )^{\frac{3}{2}   }  } = 0
\end{equation}
To complete the computation we express these answers in terms of the conserved charges.  
From the leading orders of $p$, $J_2$ and $J$ respectively, we find
\begin{eqnarray}
\tilde{C} & = & \sin\left( \frac{p}{2} \right) \qquad
a^2  =  \frac{J_2^2}{J_2^2 + \frac{\lambda}{\pi^2} r_i^2 \sin^2\left( \frac{p}{2} \right)} \\
f & = &  e^{-K} \frac{\sqrt{\lambda}}{\pi} r_i \frac{\sin^2\left( \frac{p}{2} \right)}{\sqrt{J_2^2 + \frac{\lambda}{\pi^2} r_i^2 \sin^4}} \nonumber \\
K & = & \frac{J_2^2 + \frac{\lambda}{\pi^2} r_i^2 \sin^2\left( \frac{p}{2}\right)}{J_2^2 + \frac{\lambda}{\pi^2} r_i^2 \sin^4\left( \frac{p}{2} \right)}\left( \frac{J_0 - V }{\sqrt{J_2^2 + \frac{\lambda}{\pi^2} r_i^2 \sin^2\left( \frac{p}{2} \right)}} + 1    \right) \sin^2\left( \frac{p}{2} \right).  
\end{eqnarray}
In the end, the final result is
\begin{equation}
E - J -\sqrt{J_2^2 + r_i^2 \frac{\lambda}{\pi^2} \sin^2\left( \frac{p}{2} \right)} = - 4 e^{-2 K}\frac{\frac{\sqrt{\lambda}}{\pi } r_i \sin^4\left( \frac{p}{2} \right)}{\sqrt{\frac{J_2^2}{\frac{\lambda}{\pi^2} r_i^2} + \sin^2\left( \frac{p}{2}\right) }} + o(f^4)
\end{equation}
Once again these finite size corrections are related by a simple rescaling of the 't Hooft coupling, to the result obtained
in \cite{FiniteSizeDyonic}, again exactly as predicted in \cite{Koch:2016jnm}.
In Appendix \ref{DyonicMagDiff} results for dyonic giant magnons stretched between different edges are given.  

\section{Conclusion}

We have studied a subspace of the complete Hilbert space of ${\cal N}=4$ super Yang-Mills theory. 
The subspace comprises small deformations of an LLM geometry\cite{LLM} specified by a boundary condition 
that is a set of black rings on the LLM plane.
The operators corresponding to an LLM background with a closed string excitation have a bare dimension of order $N^2$.
The large $N$ limit of correlators of these heavy operators receives contributions from non-planar diagrams even for the
leading large $N$ dynamics.
Thus, we are considering a genuinely different set up to the usual planar limit.
The fluctuations propagating on this geometry are closed strings.
When projected to the LLM plane, the closed strings are polygons with all corners lying on the outer edge of a single ring.
Our interest in these fluctuations is because the weak coupling analysis of \cite{Koch:2016jnm} shows that the net 
effect of summing the huge set of non-planar diagrams, is a simple rescaling of the 't Hooft coupling.
This immediately predicts the anomalous dimensions of these operators in terms of the
corresponding dimensions computed in the planar limit and it implies that the dynamics of this subsector is integrable.

We have carried out some highly nontrivial checks of the proposal of \cite{Koch:2016jnm}.
Using the $su(2|2)^2$ symmetry enjoyed by the subspace we consider we have determined the two magnon 
$S$-matrix and have demonstrated that it agrees up to two loops with a weak coupling computation performed in the CFT.
We have also computed the first finite size corrections to both the magnon and the dyonic magnon.
This was achieved by constructing solutions to the Nambu-Goto action that carry finite angular momentum.
These computations, which again show that the net affect of the background is the scaling of the 't Hooft coupling 
predicted in \cite{Koch:2016jnm} constitute a strong coupling check of our proposal.
These corrections are sensitive to the overall phase of the $S$-matrix which is not determined by the $su(2|2)^2$
symmetry of the theory, so this certainly appears to be a non-trivial test.

The enormous combinatoric complications of summing planar diagrams has been handled using the
representation theory methods that have been developed for heavy operators.
The final answer is remarkably simple and it suggests that there may be many other large $N$ but non-planar
subsectors of the theory that are worth exploring.

Perhaps the most significant implication of our results is the existence of other integrable sectors of ${\cal N}=4$ super
Yang-Mills theory, besides the planar limit. 
This deserves further exploration, since integrability gives us a window into the strong coupling dynamics of these 
subsectors.
It would be convincing if one could find other signatures of integrability, for example exact scattering soliton solutions
of the classical string theory which would be dual to magnon scattering.
We leave these interesting questions for the future.

{\vskip 0.25cm}
\noindent
\begin{centerline} 
{\bf Acknowledgements}
\end{centerline} 

This work is supported by the South African Research Chairs
Initiative of the Department of Science and Technology and National Research Foundation
as well as funds received from the National Institute for Theoretical Physics (NITheP).
HJR is supported by a Claude Leon Foundation postdoctoral fellowship.

\appendix

\section{Explicit Two-loop expressions}\label{klunky}

The following expressions are relevant for our two-loop computation
\begin{eqnarray}
C_1 & = & \frac{C_N}{C_D} \nonumber \\ 
C_N & = & -e^{i \beta_2}
      r_0 \left(e^{i p_2} (e^{i p_2} (r_0 - e^{i p_1} r_1) - r_2) (r_1 + 
          e^{i (p_1 + p_2)} r_2)^2 \right. \nonumber \\
		&   &			- 
       A e^{i p_1} (e^{i (p_1 + p_2)} r_1 + r_2)^2 (-e^{i p_1} r_0 + r_1 + 
          e^{i (p_1 + p_2)} r_2) \nonumber \\
		&   &			\left. + 
       B e^{i p_1} (e^{i (p_1 + p_2)} r_1 + r_2)^2 (-e^{i p_1} r_0 + r_1 + 
          e^{i (p_1 + p_2)} r_2) \right)  \nonumber \\ 
C_D & = & (e^{i (p_1 + p_2)} r_1 + r_2) (r_1 + 
       e^{i (p_1 + p_2)} r_2) (-2 e^{i p_1} r_0 + r_1 + 
       e^{i (p_1 + p_2)} r_2) \nonumber \\
C_2 & = & -C_1 \nonumber
\end{eqnarray}
		\begin{eqnarray}
N_B & = & e^{7 i p_1 + 4 i p_2} r_0^3 r_1^4 + 
      4 e^{3 i (2 p_1 + p_2)} r_0^3 r_1^3 r_2 - 
      2 e^{7 i p_1 + 5 i p_2} r_0^2 r_1^4 r_2 + 
      e^{3 i p_2} r_0 r_1^3 (r_0^2 - 2 r_1^2) r_2 + 
      6 e^{5 i p_1 + 2 i p_2} r_0^3 r_1^2 r_2^2 \nonumber \\
			& & + e^{7 i p_1 + 6 i p_2} r_0 r_1^4 r_2^2 + 
      e^{6 i (p_1 + p_2)} r_1^3 (r_0^2 - r_1^2) r_2^2 + 
      e^{2 i p_2} r_1^3 (r_0^2 + r_1^2) r_2^2 - 
      e^{i (p_1 + 2 p_2)} r_0 r_1^2 (3 r_0^2 + 4 r_1^2) r_2^2 \nonumber \\
			& & + 
      4 e^{i (4 p_1 + p_2)} r_0^3 r_1 r_2^3 - 
      e^{6 i p_1 + 7 i p_2} r_0 r_1^3 r_2^3 - e^{i p_2} r_0 r_1^3 r_2^3 + 
      e^{3i p_1} r_0^3 r_2^4 - 2 e^{2 i p_1} r_0^2 r_1 r_2^4 + 
      e^{i p_1} r_0 r_1^2 r_2^4 \nonumber \\
			& & + 
      e^{4 i p_1 + 7 i p_2} r_0 r_1 r_2^3 (r_0^2 - 2 r_2^2) + 
      e^{i (p_1 + p_2)} r_1^2 r_2^3 (r_0^2 - r_2^2) + 
      e^{5 i p_1 + 7 i p_2} r_1^2 r_2^3 (r_0^2 + r_2^2) \nonumber \\
			& & - 
      2 e^{i (3 p_1 + p_2)} r_0^2 r_2^3 (r_0^2 + 3 r_1^2 + r_2^2) - 
      2 e^{6 i p_1 + 4 i p_2} r_0^2 r_1^3 (r_0^2 + r_1^2 + 3 r_2^2) \nonumber \\ 
			& &  - 
      2 e^{2 i (2 p_1 + p_2)} r_0^2 r_1 r_2^2 (3 r_0^2 + 4 r_1^2 + 3 r_2^2) - 
      e^{5 i p_1 + 6 i p_2} r_0 r_1^2 r_2^2 (3 r_0^2 + 4 r_2^2) \nonumber \\ 
			& & - 
      2 e^{5 i p_1 + 3 i p_2} r_0^2 r_1^2 r_2 (3 r_0^2 + 3 r_1^2 + 4 r_2^2) + 
      e^{6 i p_1 + 5 i p_2} r_0 r_1^3 r_2 (r_0^2 + 3 (r_1^2 + r_2^2)) \nonumber \\
			& & + 
      e^{i (2 p_1 + p_2)} r_0 r_1 r_2^3 (r_0^2 + 3 (r_1^2 + r_2^2)) + 
      3 e^{2 i p_1 + 5 i p_2}
        r_0 r_1 r_2 (-4 r_1^2 r_2^2 + r_0^2 (r_1^2 + r_2^2)) \nonumber \\ 
				& & + 
      e^{4 i p_1 + 3 i p_2}
        r_0 r_1 r_2 (3 r_1^4 + 14 r_1^2 r_2^2 + 3 r_2^4 + 
         3 r_0^2 (r_1^2 + r_2^2)) \nonumber \\ & &  - 
      e^{2 i p_1 + _3 i p_2}
        r_0 r_1 r_2 (4 r_1^4 + 3 r_1^2 r_2^2 + 6 r_0^2 (r_1^2 + r_2^2))  - 
      e^{4 i p_1 + 5 i p_2}
        r_0 r_1 r_2 (3 r_1^2 r_2^2 + 4 r_2^4 + 6 r_0^2 (r_1^2 + r_2^2)) \nonumber \\
				& & + 
      e^{3 i (p_1 + 2 p_2)}
        r_0 r_2^2 (-8 r_1^2 r_2^2 + r_0^2 (3 r_1^2 + r_2^2)) + 
      e^{3 i p_1 + 2 i p_2}
        r_0 r_2^2 (4 r_1^4 + 10 r_1^2 r_2^2 + r_2^4 + 
         r_0^2 (3 r_1^2 + r_2^2))  \nonumber \\
				& & 
				+ 
      e^{2 i (p_1 + p_2)}
        r_1 r_2^2 (4 r_0^4 + r_0^2 r_1^2 - r_2^2 (4 r_1^2 + r_2^2)) + 
      e^{i(p_1 + 3 p_2)}
        r_1^2 r_2 (-2 r_0^4 + r_1^4 + 4 r_1^2 r_2^2 + 
         r_0^2 (6 r_1^2 + r_2^2)) \nonumber \\
				& & + 
      e^{i (p_1 + 4 p_2)}
        r_0 r_1^2 (-8 r_1^2 r_2^2 + r_0^2 (r_1^2 + 3 r_2^2)) + 
      e^{5 i p_1 + 4 i p_2}
        r_0 r_1^2 (r_1^4 + 10 r_1^2 r_2^2 + 4 r_2^4 + 
         r_0^2 (r_1^2 + 3 r_2^2)) \nonumber \\ 
				& & + 
      e^{5 i (p_1 + p_2)}
        r_1^2 r_2 (4 r_0^4 + r_0^2 r_2^2 - r_1^2 (r_1^2 + 4 r_2^2)) + 
      e^{4 i p_1 + 6 i p_2}
        r_1 r_2^2 (-2 r_0^4 + 4 r_1^2 r_2^2 + r_2^4 + 
         r_0^2 (r_1^2 + 6 r_2^2)) \nonumber \\
				& & + 
      e^{3 i (p_1 + p_2)}
        r_2 (-6 r_1^4 r_2^2 - 4 r_1^2 r_2^4 + 4 r_0^4 (2 r_1^2 + r_2^2) + 
         r_0^2 (r_1^4 - 4 r_1^2 r_2^2 + r_2^4)) \nonumber \\
				& & + 
      e^{4 i (p_1 + p_2)}
        r_1 (-4 r_1^4 r_2^2 - 6 r_1^2 r_2^4 + 4 r_0^4 (r_1^2 + 2 r_2^2) + 
         r_0^2 (r_1^4 - 4 r_1^2 r_2^2 + r_2^4)) \nonumber \\
				& & - 
      e^{3 i p_1 + 4 i p_2}
        r_0 ((r_1^2 + r_2^2)^3 + 3 r_0^2 (r_1^4 + 4 r_1^2 r_2^2 + r_2^4)) + 
      e^{2 i (p_1 + 2 p_2)} (4 r_1^5 r_2^2 + 6 r_1^3 r_2^4 - 
         2 r_0^4 (r_1^3 + 2 r_1 r_2^2) \nonumber \\
				& & + 
         r_0^2 r_1 (r_1^4 + 14 r_1^2 r_2^2 + r_2^4)) + 
      e^{3 i p_1 + 
        5 i p_2} (6 r_1^4 r_2^3 + 4 r_1^2 r_2^5 - 
         2 r_0^4 (2 r_1^2 r_2 + r_2^3) \nonumber \\
				& &  + 
         r_0^2 r_2 (r_1^4 + 14 r_1^2 r_2^2 + r_2^4)) 
\end{eqnarray}

\begin{eqnarray}
N_A & = & e^{4 i (2 p_1 + p_2)} r_0^3 r_1^4 + 4 e^{7 i p_1 + 3 i p_2} r_0^3 r_1^3 r_2 - 
  3 e^{8 i p_1 + 5 i p_2} r_0^2 r_1^4 r_2 + e^{3 i p_2} r_0^2 r_1^4 r_2 + 
  6 e^{2 i (3 p_1 + p_2)} r_0^3 r_1^2 r_2^2 \nonumber \\
	& & + 
  3 e^{8 i p_1 + 6 i p_2} r_0 r_1^4 r_2^2 - 2 e^{2 i p_2} r_0 r_1^4 r_2^2 + 
  4 e^{i (5 p_1 + p_2)} r_0^3 r_1 r_2^3 - e^{i (p_1 + p_2)} r_0 r_1^3 r_2^3 - 
  e^{7 i (p_1 + p_2)} r_0 r_1^3 r_2^3 \nonumber \\
	& & - e^{8 i p_1 + 7 i p_2} r_1^4 r_2^3 + 
  e^{i p_2} r_1^4 r_2^3 + e^{4 i p_1} r_0^3 r_2^4 - 
  3 e^{3 i p_1} r_0^2 r_1 r_2^4 + e^{5 i p_1 + 8 i p_2} r_0^2 r_1 r_2^4 + 
  3 e^{2 i p_1} r_0 r_1^2 r_2^4 \nonumber \\
	& &  - 2 e^{6 i p_1 + 8 i p_2} r_0 r_1^2 r_2^4 - 
  e^{i p_1} r_1^3 r_2^4 + e^{7 i p_1 + 8 i p_2} r_1^3 r_2^4 - 
  e^{6 i (p_1 + p_2)} r_0 r_1^2 r_2^2 (3 r_0^2 + 6 r_1^2 - 2 r_2^2) \nonumber \\
	& &  + 
  e^{4 i p_1 + 7 i p_2} r_0^2 r_2^3 (4 r_1^2 + r_2^2) + 
  e^{6 i p_1 + 7 i p_2} r_1^2 r_2^3 (5 r_0^2 + 4 r_1^2 + r_2^2) + 
  2 e^{3 i (p_1 + 2 p_2)} r_0^2 r_1 r_2^2 (3 r_1^2 + 2 r_2^2) \nonumber \\
	& &  - 
  e^{i (2 p_1 + p_2)} r_1^2 r_2^3 (3 r_0^2 + 3 r_1^2 + 2 r_2^2) - 
  e^{5 i p_1 + 7 i p_2} r_0 r_1 r_2^3 (3 r_0^2 + 8 r_1^2 + 2 r_2^2) \nonumber \\
	& & + 
  2 e^{2 i p_1 + 5 i p_2} r_0^2 r_1^2 r_2 (2 r_1^2 + 3 r_2^2) - 
  e^{7 i p_1 + 6 i p_2} r_1^3 r_2^2 (3 r_0^2 + 2 r_1^2 + 3 r_2^2) \nonumber \\
	& &  - 
  e^{i (4 p_1 + p_2)} r_0^2 r_2^3 (2 r_0^2 + 10 r_1^2 + 3 r_2^2) + 
  e^{i (p_1 + 4 p_2)} r_0^2 r_1^3 (r_1^2 + 4 r_2^2) + 
  e^{i (p_1 + 2 p_2)} r_1^3 r_2^2 (5 r_0^2 + r_1^2 + 4 r_2^2) \nonumber \\
	& &  - 
  2 e^{5 i p_1 + 2 i p_2} r_0^2 r_1 r_2^2 (3 r_0^2 + 7 r_1^2 + 5 r_2^2) + 
  e^{i (3 p_1 + p_2)} r_0 r_1 r_2^3 (5 r_0^2 + 9 r_1^2 + 5 r_2^2) \nonumber \\
	& &  - 
  e^{2 i (p_1 + p_2)} r_0 r_1^2 r_2^2 (3 r_0^2 - 2 r_1^2 + 6 r_2^2) - 
  2 e^{3 i (2 p_1 + p_2)} r_0^2 r_1^2 r_2 (3 r_0^2 + 5 r_1^2 + 7 r_2^2) \nonumber \\
	& &  - 
  e^{i (p_1 + 3 p_2)} r_0 r_1^3 r_2 (3 r_0^2 + 2 r_1^2 + 8 r_2^2) + 
  e^{7 i p_1 + 5 i p_2} r_0 r_1^3 r_2 (5 r_0^2 + 5 r_1^2 + 9 r_2^2) \nonumber \\
	& &  - 
  e^{7 i p_1 + 4 i p_2} r_0^2 r_1^3 (2 r_0^2 + 3 r_1^2 + 10 r_2^2) + 
  3 e^{5 i p_1 + 3 i p_2}
    r_0 r_1 r_2 (r_1^2 + r_2^2) (5 r_0^2 + 3 (r_1^2 + r_2^2)) \nonumber \\
	& &  - 
  e^{5 i (p_1 + p_2)}
    r_0 r_1 r_2 (6 r_1^4 + 3 r_1^2 r_2^2 - 2 r_2^4 + 6 r_0^2 (r_1^2 + r_2^2)) \nonumber \\
	& &  - 
  e^{3 i (p_1 + p_2)}
    r_0 r_1 r_2 (-2 r_1^4 + 3 r_1^2 r_2^2 + 6 r_2^4 + 6 r_0^2 (r_1^2 + r_2^2)) \nonumber \\
	& &   -
   e^{3 i p_1 + 5 i p_2}
    r_0 r_1 r_2 (8 r_1^4 + 12 r_1^2 r_2^2 + 8 r_2^4 + 9 r_0^2 (r_1^2 + r_2^2)) \nonumber \\
	& & +
   e^{2 i (2 p_1 + p_2)}
    r_0 r_2^2 (12 r_1^4 + 14 r_1^2 r_2^2 + 3 r_2^4 + 
     5 r_0^2 (3 r_1^2 + r_2^2)) \nonumber \\
	& &  + 
  e^{6 i p_1 + 4 i p_2}
    r_0 r_1^2 (3 r_1^4 + 14 r_1^2 r_2^2 + 12 r_2^4 + 
     5 r_0^2 (r_1^2 + 3 r_2^2)) \nonumber \\
	& &  - 
  e^{3 i p_1 + 2 i p_2}
    r_1 r_2^2 (4 r_1^4 + 5 r_1^2 r_2^2 + 2 r_2^4 + 
     r_0^2 (11 r_1^2 + 4 r_2^2)) \nonumber \\
	& &  + 
  e^{5 i p_1 + 6 i p_2}
    r_1 r_2^2 (2 r_0^4 + 6 r_1^4 + 4 r_1^2 r_2^2 + r_2^4 + 
     r_0^2 (17 r_1^2 + 6 r_2^2)) \nonumber \\
	& &  - 
  e^{6 i p_1 + 5 i p_2}
    r_1^2 r_2 (2 r_1^4 + 5 r_1^2 r_2^2 + 4 r_2^4 + 
     r_0^2 (4 r_1^2 + 11 r_2^2)) \nonumber \\
	& &  + 
  e^{2 i p_1 + 3 i p_2}
    r_1^2 r_2 (2 r_0^4 + r_1^4 + 4 r_1^2 r_2^2 + 6 r_2^4 + 
     r_0^2 (6 r_1^2 + 17 r_2^2)) \nonumber \\
	& &  - 
  e^{4 i (p_1 + p_2)}
    r_0 ((r_1^2 + r_2^2)^3 + 3 r_0^2 (r_1^4 + 4 r_1^2 r_2^2 + r_2^4)) \nonumber \\
	& &  - 
  e^{4 i p_1 + 6 i p_2}
    r_0 r_2^2 (3 r_0^2 (3 r_1^2 + r_2^2) + 
     2 (6 r_1^4 + 4 r_1^2 r_2^2 + r_2^4)) \nonumber \\
	& &  - 
  e^{4 i p_1 + 3 i p_2}
    r_2 (3 r_1^6 + 6 r_1^4 r_2^2 + 5 r_1^2 r_2^4 + r_2^6 + 
     r_0^2 (11 r_1^4 + 16 r_1^2 r_2^2 + 3 r_2^4)) \nonumber \\
	& &  + 
  e^{4 i p_1 + 5 i p_2}
    r_2 (4 r_1^6 + 6 r_1^4 r_2^2 + 4 r_1^2 r_2^4 + r_2^6 + 
     2 r_0^4 (2 r_1^2 + r_2^2) + 
     r_0^2 (17 r_1^4 + 22 r_1^2 r_2^2 + 5 r_2^4)) \nonumber \\
	& &  - 
  e^{2 i (p_1 + 2 p_2)}
    r_0 r_1^2 (3 r_0^2 (r_1^2 + 3 r_2^2) + 
     2 (r_1^4 + 4 r_1^2 r_2^2 + 6 r_2^4)) \nonumber \\
	& &  - 
  e^{5 i p_1 + 4 i p_2}
    r_1 (r_1^6 + 5 r_1^4 r_2^2 + 6 r_1^2 r_2^4 + 3 r_2^6 + 
     r_0^2 (3 r_1^4 + 16 r_1^2 r_2^2 + 11 r_2^4)) \nonumber \\
	& &  + 
  e^{3 i p_1 + 4 i p_2}
    r_1 (r_1^6 + 4 r_1^4 r_2^2 + 6 r_1^2 r_2^4 + 4 r_2^6 + 
     2 r_0^4 (r_1^2 + 2 r_2^2) + r_0^2 (5 r_1^4 + 22 r_1^2 r_2^2 + 17 r_2^4)) 
\eea

\bea
		D_A & = & (e^{i (p_1 + p_2)} r_1 + r_2)^3 (r_1 + 
    e^{i (p_1 + p_2)} r_2) (-2 e^{i p_1} r_0 + r_1 + e^{i (p_1 + p_2)} r_2)^2\nonumber 
\end{eqnarray}

\section{Finite Size Computations}\label{evaluate}

In this section we outline the evaluation of the integrals needed to evaluate the finite size correction.  
We will evaluate the integral
\begin{equation}
I = \int_{f}^{1} \frac{\sqrt{z^2 - f^2}}{\sqrt{1 - z^2}} F(z) dz
\end{equation}
as a series expansion in $f$.  
It is sufficient to assume that $F(z)$ is a positive function, finite at $z = f$ and $z = 1$.  
Both the integrand and the limits of the integral have $f$ dependence.  
To start, eliminate the $f$-dependence from the limits of integration.  
Write
\begin{eqnarray}
I & = &  \int_{0}^{1} \frac{\sqrt{z^2 - f^2}}{\sqrt{1 - z^2}} F(z) dz  - \int_{0}^{1} dz' f^2 \frac{\sqrt{1 - z'^2}}{\sqrt{1 - f^2 z'^2}} F(f z') \nonumber \\
  & \equiv & I_1 - I_2.  
\end{eqnarray}
The second integrand can be expanded in powers of $f$ to yield a series of convergent integrals.  
The first integral, on the other hand, leads to log divergent integrals.
To capture these divergences change variables as follows
\begin{equation}
I_1 =  \int_{-\infty}^{\log(\frac{1}{f})} dy f e^y \frac{1}{\sqrt{1 - f^2 e^{2y}}} F(f e^{2y})  =  \int_{-\infty}^{\log(\frac{1}{f})} dy f e^y \tilde{F}(f e^{y}) .  
\end{equation}
Now perform a series expansion of the integrand
\begin{equation}
I_1 =  \int_{-\infty}^{\log(\frac{1}{f})} dy f e^y \sum_{i=1}^\infty \tilde{F}^{(i)} (f e^y)^i 
\end{equation}
and terminate the expansion at a given order
\begin{equation}
\tilde{I}_1^{(n)} = \int_{-\infty}^{\log(\frac{1}{f})} dy f e^y \sum_{i=1}^n \tilde{F}^{(i)} (f e^y)^i
\end{equation}
The upper limit of the integral carries an $f$ dependence.  
To capture this dependence compute
\begin{equation}
I_3^{(n)} =  \int_{0}^{1} \frac{1}{\sqrt{1 - z^2}} F(z) dz - \int_{0}^{1} dz \sum_{i=1}^n \tilde{F}^{(i)} z^i.  
\end{equation}
The expansion of the integral $I_3$ yields a sequence of convergent integrals.  
Putting everything together we can perform a series expansion of the original integral as 
\begin{equation}
I \approx I^{(n)} =  I_{3}^{(n)} + I_{1}^{n} - I_2^{(n)}
\end{equation}

\subsection{$I = p$}

We list the integrals needed to reproduce the final results in the main text.  
Working to second order in $f$ we find
\begin{eqnarray}
\tilde{I}_1 & = & \frac{1}{2}\tilde{C}\sqrt{1 - \tilde{C}^2} + \frac{1}{4}\left(\pi - 1 - \log(4) + 2 \log(f) \right)f^2  + o(f^4) \nonumber \\
 I_2 & = & \frac{\pi}{4} \sqrt{1 - \tilde{C}^2} \tilde{C} f^2 + o(f^4) \nonumber \\
I_3 & = & \frac{1}{2}\left( - \tilde{C}\sqrt{1 - \tilde{C}^2}+ \pi + 2 i \log\left( \tilde{C} + i\sqrt{1 - \tilde{C}^2} \right)\right) - \frac{1}{2} \tilde{C}\sqrt{1 - \tilde{C}^2} \log(2) f^2  + f^2 I_4 + o(f^4) \nonumber
\end{eqnarray}
where 
\begin{equation}
I_4 = -\int_{0}^1 dz \frac{\tilde{C}\sqrt{1 - \tilde{C}^2}}{2 z \sqrt{1 - z^2}(1 - \tilde{C}^2 z^2)} \bar{V_{\phi}}(z)
\end{equation}

\subsection{$I = R_p$}

Working to second order we find
\begin{eqnarray}
\tilde{I}_1 & = & \frac{1}{2}\tilde{C}\sqrt{1 - \tilde{C}^2}r_0^2 + \frac{1}{4}\tilde{C}\sqrt{1 - \tilde{C}^2}r_0^2\left(\pi - 1 - \log(4) + 2 \log(f) \right)f^2  + o(f^4) \nonumber \\
 I_2 & = & r_0^2 \frac{\pi}{4} \sqrt{1 - \tilde{C}^2} \tilde{C} f^2 + o(f^4) \nonumber \\
I_3 & = & \frac{1}{2} \tilde{C}\sqrt{1 - \tilde{C}^2}r_0^2 + \frac{1}{2} \tilde{C}\sqrt{1 - \tilde{C}^2}r_0^2(\tilde{C}^2 - \log(2))f^2 + f^2 I_5 + o(f^4) 
\end{eqnarray}
where 
\begin{equation}
I_5 = -\int_0^1 dz \frac{\tilde{C}\sqrt{1 - \tilde{C}^2}r_0^2 }{2 z \sqrt{1 - z^2}} \bar{V}_{\phi}(z).  
\end{equation}

\section{Dyonic Magnon solutions stretched between different edges}\label{DyonicMagDiff}

Magnons stretched between different edges lead to two integrals with different limits.  
One integral runs from $r = r_m$ to $r = r_i$, while the second integral runs from $r = r_m$ to $r = r_j$.  
We can evaluate both integrals using the change of variables $r_m = r_k\sqrt{1 - \tilde{C}^2_k}$.  
The constant $\tilde{C}$ for the two integrals is different.  
To compare the values of the integrals we transform back to $r_m$ and $r_k$
\begin{equation}
\tilde{C}_k = \sqrt{1 - \frac{r_m^2}{r_k^2}}.  
\end{equation}
The conserved charges are then sums of the form
\begin{equation}
p = p_i(r_i) + p_j(r_j)
\end{equation}
which reduces to our previous expressions when $r_j = r_i$.  
Solving for the constants in terms of conserved charges is involved but tractable. 
After simplification we find the following finite size corrections
\begin{eqnarray}
& & E-J \nonumber \\
& = & \sqrt{J_2^2 + \frac{\lambda}{4\pi^2}(r_i^2 + r_j^2) - \frac{\lambda}{2\pi^2}\cos(p) } +  \nonumber \\
& & \frac{1}{16(r_i^2 + r_j^2 - 2 r_j r_j \cos(p))\sqrt{J_2^2 + \frac{\lambda}{4\pi^2}(r_i^2 + r_j^2) - \frac{\lambda}{2\pi^2}\cos(p) }}\times \left( -4(J_2^2(r_i^2 + r_j^2) + \right. \nonumber \\
& & \left. \frac{\lambda}{4 \pi^2}(r_i^4 + 3 r_i^2 r_j^2 + r_j^4)      )   + r_i r_j(8 J_2^2 + 15 \frac{\lambda}{4\pi^2}(r_i^2 + r_j^2)   )\cos(p) \right. \nonumber \\
& & \left. +   \frac{\lambda}{4 \pi^2} r_i r_j(-12 r_i r_j \cos(2p) + (r_i^2 + r_j^2)\cos(3p)  ) \right)  f^2.  
\end{eqnarray}
This expression has the correct limits when $J_2 \rightarrow 0 $ and $r_i \rightarrow r_j$.  
The constant $f$ is determined from 
\begin{equation}
J_0 - V_i - V_j = \sum_{k = i,j} \frac{\sqrt{\lambda}( -2r_k^2 + 2r_m^2 + (r_k^2 - (1-a^2)r_m^2)\log\left( \frac{16(1-a^2)(r_k^2 - r_m^2)}{f^2(r_k^2 - (1-a^2)r_m^2)}   \right)   )}{4\pi \sqrt{1 - a^2}\sqrt{r_k^2 - r_m^2}}
\end{equation}
where 
\begin{eqnarray}
r_m & = & \frac{r_i r_j \sin(p)}{\sqrt{r_i^2 + r_j^2 - 2 r_i r_j \cos(p)}} \nonumber \\
a & = & -\frac{ J_2 \sqrt{J_2^2 + \frac{\frac{\lambda}{4\pi^2}(r_i^2 - r_j^2)}{ r_i^2 + r_j^2 - 2 r_i r_j \cos(p)  }   }  }{ \sqrt{J_2^4 + \frac{\lambda^2}{16\pi^4} (r_i^2 - r_j^2)^2 +  J_2^2 \frac{\lambda}{2\pi^2} (r_i ^2 + r_j^2 - 2\frac{r_i^2 r_j^2 \sin^2(p)}{r_i^2 + r_j^2 - 2 r_i r_j \cos(p)})}    }
\end{eqnarray}

\vfill\eject
\end{document}